\documentclass[sigconf]{acmart}
\def\BibTeX{{\rm B\kern-.05em{\sc i\kern-.025em b}\kern-.08emT\kern-.1667em\lower.7ex\hbox{E}\kern-.125emX}}

\renewcommand\footnotetextcopyrightpermission[1]{} 

\def\BibTeX{{\rm B\kern-.05em{\sc i\kern-.025em b}\kern-.08em
    T\kern-.1667em\lower.7ex\hbox{E}\kern-.125emX}}
    
\usepackage{graphicx}
\graphicspath{ {images/} }
\usepackage[font=small,skip=4pt]{caption}
\usepackage[font=small,skip=0pt]{subcaption}

\usepackage{pgfplots}
\usepackage{tikz}
\usetikzlibrary{patterns}
\usepackage[utf8]{inputenc}
\usepackage[english]{babel}

\usepackage{amsmath}
\usepackage{amsthm}
\usepackage{xspace}
\usepackage{mathtools}
\usepackage{url}
\usepackage{booktabs}
\usepackage{mdframed}
\usepackage{lipsum}

\usepackage{enumitem}
\setlist[itemize]{noitemsep, nolistsep}
\setlist[enumerate]{noitemsep, nolistsep}

\usepackage{algorithm}
\usepackage{multicol}
\usepackage[noend]{algpseudocode}

\newtheorem{claim}{Claim}

\newcommand{\prot}{{\sc Rivet}}
\newcommand{\qrm}{{$\sf Quorum$}}
\newcommand{\ibft}{IBFT}
\newcommand{\tpc}{2PC}
\newcommand{\dte}{DTE}

\newcommand{\Ws}{Worker}
\newcommand{\Rs}{Reference}

\newcommand{\ws}{worker}
\newcommand{\rs}{reference}
\newcommand{\cs}{coordinator}

\newcommand{\two}{\vspace{2mm}}

\newcommand{\hgt}{{\sf height}}
\algnewcommand{\LineComment}[1]{\State \(\triangleright\) #1}

\let\emptyset\varnothing
\let\px\preccurlyeq

\newcommand{\bt}{{\bf T}}
\newcommand{\bx}{{\bf X}}
\newcommand{\bq}{{\bf Q}}

\newcommand{\bc}{{\bf C}}
\newcommand{\bh}{{\bf H}}
\newcommand{\br}{{\bf R}}
\newcommand{\bw}{{\bf W}}

\newcommand{\adv}{${\mathcal{A}}$\xspace}

\newcommand{\st}{{\sf state}}
\newcommand{\hs}{{\sf hash}}
\newcommand{\tx}{{\sf tx}}
\newcommand{\ctx}{{\sf ctx}}
\newcommand{\cm}{{\sf com}}
\newcommand{\key}{{\sf key}}

\newcommand{\ch}{\mathcal{C}}

\newcommand{\la}{\langle}
\newcommand{\ra}{\rangle}

\newcommand{\intv}{{\mathbb I}}

\definecolor{hoyemagenta}{HTML}{FF3FFF}
\definecolor{hoyegreen}{HTML}{31FF31}

\usepackage{tikz}
\usepackage{amsmath}

\usepackage{filecontents}
\usepackage[T1]{fontenc}
\setcopyright{none} 

\begin{document}

\date{}


\fancyhead{}
\def\thetitle{Efficient Cross-Shard Transaction Execution in Sharded Blockchains}
\title{\thetitle}

\author{Sourav Das}
\affiliation{University of Illinois at Urbana-Champaign}
\email{souravd2@illinois.edu}
\author{Vinith  Krishnan}
\affiliation{University of Illinois at Urbana-Champaign}
\email{vinithk2@illinois.edu}
\author{Ling Ren}
\affiliation{University of Illinois at Urbana-Champaign}
\email{renling@illinois.edu}

\date{}

\begin{abstract}
Sharding is a promising blockchain scaling solution. But it currently suffers from  high latency 
and low throughput when it comes to cross-shard 
transactions, i.e., transactions that require coordination from multiple shards. 
The root cause of these limitations arise from the use of the classic two-phase commit protocol, which involves locking assets for extended periods of time.
This paper presents \prot, a new paradigm for blockchain sharding that achieves lower latency and higher throughput for cross-shard transactions. \prot\ has a single \rs\ shard running consensus, and multiple \ws\ shards maintaining disjoint states and processing a subset of transactions in the system. 
\prot\ obviates the need for consensus within each \ws\ shard, and as a result, tolerates more failures within a shard and lowers communication overhead.
We prove the correctness and security of \prot.
An evaluation of our prototype implementation atop 50+ AWS EC2 instances demonstrates the latency and throughput improvements for cross-shard transactions of \prot\ over the baseline \tpc. 
As part of our evaluation effort and an independent contribution, 
we also propose a more realistic framework for evaluating sharded
blockchains by creating a benchmark based on real Ethereum transactions. 
\end{abstract}
\maketitle

\section{Introduction}
\label{sec:introduction}
A typical blockchain system replicates storage and computations 
among all its nodes and runs a single consensus algorithm 
involving all nodes~\cite{nakamoto2008bitcoin,wood2014ethereum}.
Such a global replication approach has limited scalability and 
throughput. 
Sharding has emerged as a promising approach to address the 
long-standing quest for blockchain scalability~\cite{luu2016secure,
al2017chainspace,kokoris2018omniledger,zamani2018rapidchain,
dang2019towards,wang2019monoxide}. 
Sharding improves scalability by partitioning different 
responsibilities and resources to different sets of nodes. 
A sharded blockchain can potentially shard its storage, communication,
and computation. 

A critical design component of a sharded ecosystem is its mechanism
to handle \emph{cross-shard transactions},
i.e., transactions that involve more than one shards. 
Cross-shard transactions are essential to sharded blockchains as they enable users to atomically interact with multiple shards;
in other words, a sharded blockchain without such support is uninteresting as it degenerates to running multiple independent blockchains. 
Popular examples of cross-shard transactions include atomic exchange of assets maintained at different shards~\cite{herlihy2018atomic}, and atomically booking a flight ticket and a hotel room where 
the two are being sold in different
shards~\cite{disttxn,train2018hotel}.

A number of prior works~\cite{luu2016secure,al2017chainspace,
kokoris2018omniledger,zamani2018rapidchain,dang2019towards,
wang2019monoxide} proposed sharding schemes under different settings. 
These protocols can linearly scale intra-shard transactions,
i.e., transactions that can be processed within a single shard,
by adding more shards to the system.
However, existing works encounter a performance bottleneck when it comes to cross-shard transactions.
All of the above works adopt the two-phase commit~(\tpc) protocol 
to execute cross-shard transactions.
While \tpc\ is the simplest and most well-known atomic commit 
protocol, it requires nodes to lock assets for an extended 
period of time, leading to higher latency and lower throughput 
for cross-shard transactions. 

\two 
\noindent
{\bf A new paradigm for sharded blockchains.}
In this paper, we aim to address the above limitation with a new framework for sharded blockchains called \prot. 
\prot\ achieves lower confirmation latency and better throughput for cross-shard transactions at a modest cost of high intra-shard latency (but not throughput). We give an overview of \prot\ below. 

\prot\ has a single \emph{\rs} shard and multiple \emph{\ws} shards.
Each shard can be both permissioned and permissionless. This 
paper focuses on the permissioned setting. In particular, we
assume that every node in a shard is aware of identities 
of other nodes of its own and the \rs\ shard.\footnote{
Note that this assumption is implicit in all committee based
Byzantine Fault Tolerant consensus protocols.}
The \rs\ shard runs a consensus layer and maintains its 
own blockchain. Each \ws\ shard maintains a disjoint set of 
states in the system.
Each \ws\ shard executes blocks of transactions involving it and vouches for 
the validity of the resulting state.
It is important to note that \ws\ shards do not run consensus on these blocks -- instead, they periodically submit hash 
digests of \ws\ blocks to the \rs\ shard.
Cross-shard transactions are also submitted to the \rs\ shard by users in the system.
The \ws\ shard commitments and cross-shard transactions are then 
finalized and ordered by the consensus layer of the \rs\ shard. 
When a set of cross-shard transactions are finalized, each \ws\ shard locally 
executes the subset of these transactions that are relevant to it, 
atop the latest committed states. To do that, a \ws\ shard needs to
download the data needed by these transactions from other shards along
with accompanying proofs showing the validity of the data (under the 
latest commitments).

\prot\ offers two main advantages over the classic 2PC approach. 
Firstly, \prot\ obviates the need to run consensus within each \ws\ shard.
As a result, each \ws\ shard in \prot\ requires few 
replicas\footnote{We use the terms replica and node interchangeably 
in this paper.} and runs a simpler and cheaper (using less 
communication) protocol, compared to the \tpc\ approach.  
%
%
Secondly, \prot\ improves the confirmation latency and throughput of cross-shard transactions.
Specifically, a cross-shard transaction gets confirmed as 
soon a single \ws\ shard involved in the transaction locally 
executes it and adds it to a certified \ws\ block~(\S\ref{sec:design}).
This holds independent of the number of shards involved, and is in sharp contrast to the \tpc\ approach, where cross-shard transactions are delayed by the slowest participating shards. 
As a consequence of the lower latency, cross-shard transactions in \prot\ lock data items for a shorter amount of time (i.e., they are made available to future transactions sooner), leading to higher throughput for cross-shard transactions.
%
Not running consensus protocols in \ws\ shards also comes with a downside: intra-shard transactions are finalized only when their state commitments get included in the \rs\ chain. 
This will result in a higher latency for intra-shard transactions compared to \tpc. 

We implement \prot\ (and a \tpc\ baseline) atop open-source 
Quorum client~\cite{jpmorgan2020quorum}. 
On a side note, our implementation supports the generic smart contract execution model whereas prior sharding proposals focus on the Unspent Transaction Output~(UTXO)~\cite{nakamoto2008bitcoin}  
transaction model, which might be of independent interest to 
the reader.

\two
\noindent
{\bf An evaluation framework for sharded blockchains.}
While attempting to compare \prot\ against the \tpc\ approach, we find us (and the field of blockchain sharding) in need of a better evaluation framework. 
Currently, the evaluation methodology of existing works is ad-hoc and artificial.
In particular, most of them randomly allocate (synthetic) transactions to shards.
Clearly, such a random allocation would make the vast majority of transactions cross-shard and fail to capture the characteristics of a realistic sharded blockchain. 

In light of this, we try to characterize the Ethereum transaction history with the aim of better understanding the interactions within and across shards had we sharded Ethereum.  
We proceed to create a benchmark for sharded blockchains. 
At a high level, our benchmark represents interactions between accounts as a graph and partitions them into different shards while minimizing the amount of cross-shard transactions. Overall, we observe less than 
30\% 
cross-shard transactions among different shards as opposed 
to over 90\% cross-shard transactions arising from a random allocation of accounts to shards~\cite{zamani2018rapidchain,dang2019towards}. 
We observe that our approach partitions 
major services along with their users into different 
shards. Thus, we believe this benchmark gives a more realistic way to evaluate sharded blockchain systems (ours and 
future ones). 

\two
\noindent
{\bf Experimental Evaluation.}
We then evaluate them using 
our benchmark on a testbed of 50+ AWS EC2 instances with realistic 
network delays. Our evaluation illustrates that 
almost all cross-shard transactions in \prot\ are confirmed 
within one \ws\ block interval from its inclusion in the 
\rs\ chain.
Furthermore, \prot\ has approximately 35\% better throughput 
for cross-shard transactions in comparison to \tpc\ based design.
In addition, the vast majority ($>99\%$) of state variables 
accessed by cross-shard transactions are unlocked immediately 
in \prot\ unlike in \tpc. 

\vspace{1mm}
\noindent In summary, we make the following contributions:
\begin{itemize}
    \item We present \prot, a novel sharded system that has  
    lower confirmation latency for cross-shard transactions,
    tolerates more failures, and has better block utilization over 
    existing approaches. We supplement our claims with theoretical 
    proofs of their correctness and security. 
    \item We analyze historical Ethereum transactions to better characterize
    benefits of sharding in permissionless blockchains and use our analysis
    to create a realistic benchmark for evaluating sharded blockchains.
    \item We implement both \prot\ and \tpc\ atop an open source \qrm\ client 
    and rigorously evaluate them using our benchmark on a testbed of 50+ 
    AWS EC2 instances. Our evaluations further corroborate our design
    choices. 
\end{itemize}

\vspace{1mm}
\noindent
{\bf Paper Organization.}
We give the background in \S\ref{sec:system}. 
We describe our methodology to analyze Ethereum transaction history 
and our findings in~\S\ref{sec:case}.
We present the detailed design of \prot\ in~\S\ref{sec:design} and 
argue about its correctness and security in~\S\ref{sec:analysis}~(with 
formal proofs in Appendix~\ref{apx:security}). \S\ref{sec:eval} 
describes our prototype implementation of \prot\ and \tpc\ as well 
as experimental results. 
We describe related work in~\S\ref{sec:related} and end with a 
discussion on future research directions in~\S\ref{sec:discussion}.

\section{Preliminaries and Notation}
\label{sec:system}
\noindent
{\bf Sharded Blockchains.}
Sharding is a prominent approach used to improve the performance and scalability of current blockchain protocols.
The main idea is to split the overheads of processing transactions among multiple smaller groups of nodes. These groups, also called shards, work in parallel to maximize performance (involving transaction processing and state update) while requiring less communication with fewer other nodes.
This allows the system to scale to a large number of participating nodes.

To begin with, the participating nodes must be partitioned into different 
groups in such a manner that no shard is overwhelmed by too many 
malicious nodes. This is typically done by partitioning the nodes 
randomly into approximately equal sized
groups~\cite{luu2016secure,kokoris2018omniledger,
zamani2018rapidchain}.
The size of the groups are set so that once partitioned, the fraction of faulty nodes in each shard remains below a certain threshold.

The state of the blockchain is then  partitioned amongst different 
shards, i.e. a disjoint set of accounts are assigned to each shard 
so that nodes in one shard only process transactions associated 
with those accounts. 
Sharding is expected to improve performance because, hopefully, 
most transactions are ``local'' to a single shard and only 
require the  participation of replicas maintaining that shard.
We call these transactions {\em intra}-shard transactions. 



\two
\noindent
{\bf Cross-shard transactions and \tpc.}
\label{sub:sharding}
Transactions that
involve multiple shards are 
called {\em cross}-shard transactions.
Execution of cross-shard transactions require some coordination mechanism among 
the participating shards. 
Most existing sharding schemes use 2PC to atomically execute 
cross-shard transactions.
Moreover, they primarily focus on UTXO based model where each transaction uses unspent tokens as inputs to create a new transaction with fresh unspent outputs. 
We use an example to illustrate how such a sharding system works.  

Say a user creates a cross-shard transaction that takes two unspent tokens, $u_1$ on a shard $X_1$ and $u_2$ on a second shard $X_2$, and moves them to a third shard $X_3$.
The creator of the transaction,
referred to as the client, broadcasts this transaction to the two 
input shards $X_1$ and $X_2$. On receiving this transaction, 
the two input shards first validate it, i.e., check whether the tokens are indeed unspent;
if so, an input shard locks the input and produces an approval 
certificate (e.g., signed by sufficiently many replicas within the shard) confirming the validity of the input. 
On the contrary, say one of the inputs is invalid, e.g., the associated 
token has already been spent, the corresponding input shard produces
a rejection certificate indicating the invalidity of the input. 
This is the first (locking) phase in classic 2PC. 
Note that once an input is locked, no future 
transaction can use the input until it is unlocked.

The client waits for the certificates from all input shards, and if
all input shards unanimously approve the transaction, it sends 
the transaction along with all the certificates to the output 
shard(s) ($X_3$ in our example). 
On receiving the cross-shard transaction and the unanimous approval certificates,
the output shard adds the desired token to the appropriate account and sends 
a confirmation certificate to the client. Alternatively, if any 
of the input shards reject its input, every output shard rejects 
the transaction. The client also forwards the approval or rejection
certificates to every input shard.
An input shard marks the input as spent if there are unanimous approval certificates,
or else unlocks the input for future transactions. 
This 
is the second phase of the standard \tpc\ protocol.

\two
\noindent
{\bf Smart contracts.}
Smart contracts are programs consisting of a set of functions that are identified by unique addresses. Each smart
contract maintains its state, a set of disjoint key-value pairs, that 
can be modified according to the program logic of the contract. 
Smart contracts are created by sending transactions containing
its code. Upon creation, users can 
invoke functions in them by sending transactions to the contract 
address. Functions of smart contracts can also be invoked 
by other smart contracts. 

A transaction invokes a function by specifying the appropriate contract address, the function, 
and the required arguments to the function. 
On receiving a transaction, 
the proposer of a block validates the transaction before including it in its 
proposal.
Once included in a proposal, transactions are executed atop 
some initial state, and its execution results in a new state. 
The state transition is deterministic and is denoted by the 
function $\Pi$. 
Specifically, if a transaction $\tx$ is executed on top of an initial state $\st$, then the resulting state is $\st'=\Pi(\st,\tx)$. 
Sometimes, we overload the notation to apply the transition function 
$\Pi$ on an ordered list of transactions, which should be interpreted as executing the transactions in the ordered list one by one.

\section{A Benchmark for Sharded Blockchains}
\label{sec:case}
%

Prior sharding works partition the state among shards in a 
uniformly random manner. Clearly, such a random partitioning 
does not capture a realistic workload for sharded blockchains.
In particular, it will result in a dominant fraction of
cross-shard transactions~\cite{zamani2018rapidchain,
dang2019towards}. Intuitively, one would expect lot more
structure/patterns between cross-shard and intra-shard
transactions than the simulated transactions with random 
access pattern. Consequently, evaluation results from these 
contrived benchmarks may significantly depart from reality 
and fail to 
accurately reflect the performance of sharded blockchains. 
In this section, we seek to create a benchmark suitable for sharded blockchains by intelligently partitioning the workload of Ethereum, which is a leading blockchain supporting general computation in the real world. 

To this end, we partition the Ethereum state in such a way that cross-shard interactions are minimized.
We analyze our results, and observe that major ''services'' are assigned to different shards. Moreover, many other accounts interact with one major service frequently and they are assigned to the same shard as that service. 
We believe this will be close to the ecosystem of a realistic sharded blockchain and the benchmark created this way is a good candidate for evaluating sharded blockchains in this paper as well as future works.

\subsection{Methodology}
\label{sub:method}

We take four thousand different blocks 
starting approximately at the 7.3 million'th block. 
We represent accounts and transactions interaction with them as an undirected graph. Each account is a 
vertex.
Edge weights denote the number of transactions that involve the corresponding two accounts.
For every transaction that involves accounts $u$ and $v$ both, the edge weight of $(u,v)$ is incremented by one.
If a transaction involves more than two accounts, it contributes one unit of weight to all edges in the clique formed by these accounts.


Our partitioning scheme is inspired by techniques used in distributed database partitioning. The connection will be explored in section~\S\ref{sec:related}.
As mentioned, we hope to partition the accounts into a number of disjoint shards and minimize the number of cross-shard transactions.
But, a blunt partitioning approach will simply put all accounts in a single shard and eliminate cross-shard transactions.
Thus, we need additional constraints to avoid the above trivial partition results.
To this end, we require the partition to be more or less \emph{balanced} in terms of activities.
In particular, we will assign every vertex four different weights:
(1) the account's storage size (measured in bytes)
(2) the total degree of 
the vertex, i.e., the total number of transactions that
access the vertex, 
(3) the total amount of computation (measured in gas) used by the transactions accessing the account, and 
(4) the total size of the transactions accessing the account. 
These four weights measure the storage, frequency of involvement, computation, and communication associated with an account, respectively.

We then seek to partition the graph into non-overlapping shards such that the total weight of the cross-shard edges are minimized (i.e. a min-cut) and all shards are balanced within a constant factor in terms of each of the four aggregated weights. 
For each of the four metrics (storage, number of involvement, computation, and communication), the aggregated weight of a shard is the sum of the corresponding weights 
of the vertices assigned to the shard.
We use the Metis tool~\cite{karypis1998software} -- a heuristic tool for constrained $k$-way graph partitioning -- to perform the partitioning for different values of $k$.

\begin{figure}[t]
    \captionsetup[subfigure]{aboveskip=-4pt,belowskip=-4pt}
    \centering
    \begin{subfigure}{0.60\linewidth}
    \centering
    \includegraphics[width=\linewidth]{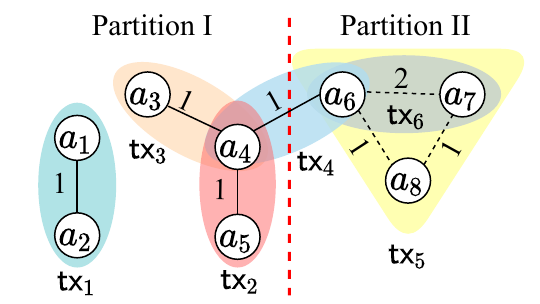} 
    \end{subfigure}    
    \begin{subfigure}{0.30\linewidth}
    \centering
    \includegraphics[width=\linewidth]{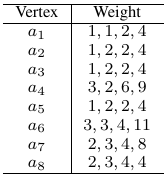} 
    \end{subfigure} 
    \caption{Sample graph with eight accounts $\{a_1,\ldots,a_8\}$ 
    and six transactions $\{\tx_1,\ldots,\tx_6\}$ marked with
    regions of different colors. Vertex weights is a four element 
    tuple $($degree, state size, gas usage, transaction size$)$.}
    \label{fig:tx-graph}
\end{figure}



%
Figure~\ref{fig:tx-graph} illustrates our approach on a sample
state with eight accounts $\{a_1,\ldots,a_8\}$ and six 
transactions $\{\tx_1,\ldots,\tx_6\}$ indicated by the 
colored regions. Accounts accessed by the transaction are enclosed
by their respective regions. For example, transaction $\tx_5$
accesses $a_6,a_7$ and $a_8$. Account $a_4$ is accessed by 
$\tx_2,\tx_3$ and $\tx_4$. Edge weights between a pair 
of nodes represent the number of times the pair has been accessed
by common transactions. For example, edge $(a_6,a_7)$ has a
weight of $2$ as the pair has been accessed by both $\tx_5$
and $\tx_6$. Also, the aggregated weights of accounts in 
both partition are balanced.

\subsection{Partitioning Results and Analysis}
\label{sub:findings}
%

\begin{figure}[t!]
    \centering
    \pgfplotsset{footnotesize,height=4.5cm, width=0.95\linewidth}
    \begin{tikzpicture}
    \begin{axis}[
        legend pos=north east,
        legend columns=2,
        xtick=\empty,
        extra x ticks={0,10,20,30,40,50,60,70},
        xlabel={Number of worker shards $k$},
        ylabel= {Cross-shard fraction},
        ymax=50,
        ] 
		\addplot [blue, mark=x] table [x=shard_num, y=avg_val, col sep=comma] {data/ctxFrac.csv};
		\addplot [red] table [x=shard_num, y=desired_load, col sep=comma] {data/ctxFrac.csv};
        \addlegendentry{Observed }
        \addlegendentry{Desired, $1/(1+k)$}
    \end{axis}
    \end{tikzpicture}
    \caption{Average fraction of transactions that are cross-shard with varying
    number of shards, $k$, evaluated by partitioning a trace of approximately 750 thousand historical Ethereum transactions. The average is taken over 5 different ranges of 1000 blocks each.}
    \label{fig:cross-frac}
\end{figure}
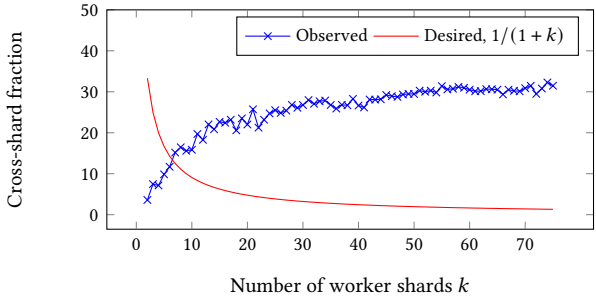

The first decision we need to make in creating a benchmark is how many shards we should have in total. 
We will employ a heuristic discussed below. 
We have already discussed that we aim to make every shard obtained by partitioning have roughly the same number of transactions for processing, resulting in a balanced workload. 
Following the same principle, we would also like to make the number of cross-shard transactions roughly the same as the number of intra-shard transactions per shard, again resulting in a balanced workload between \ws\ shards and \rs\ shard (or \cs\ shard).  
This means, we should try to make the fraction of cross-shard transactions roughly the reciprocal of the total number of shards, i.e., $1/(k+1)$ where $k$ is the number of \ws\ shards. 
Figure~\ref{fig:cross-frac} illustrates the  fraction of cross-shard transactions obtained and the desired target of $1/(k+1)$ as a function of the number of shards. 
Naturally, the fraction of cross-shard transactions increases with the number of shards (in the extreme case of a single \ws\ shard, all transactions are intra-shard), while the desired fractions decreases. The two curves intersect roughly when the number of worker shards is 6. This is the number of worker shards we will use in our experiments.

%
%

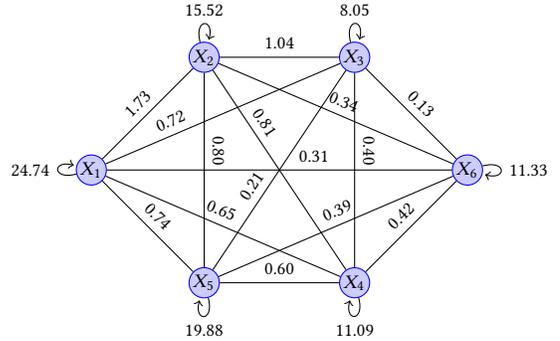
\begin{figure}[tb]
    \captionsetup[subfigure]{aboveskip=-4pt,belowskip=-4pt}
    \centering
    \begin{tikzpicture}
    [scale=0.5, auto, node distance=auto,
      node_style/.style={circle,draw=blue,fill=blue!20!,inner sep=0.5pt,minimum size=1pt,font=\footnotesize},
      edge_style/.style={draw=black,font=\footnotesize}]
        \node[node_style] (v2) at (-2,3) {$X_2$};
        \node[node_style] (v3) at (2,3) {$X_3$};
        \node[node_style] (v6) at (5,0) {$X_6$};
        \node[node_style] (v4) at (2,-3) {$X_4$};
        \node[node_style] (v5) at (-2,-3) {$X_5$};
        \node[node_style] (v1) at (-5,0) {$X_1$};
        \path (v1) edge [loop left,font=\footnotesize] node {24.74} (v1);
        \draw[edge_style]  (v1) edge node[sloped,above]{1.73} (v2);
        \draw[edge_style]  (v1) edge node[pos=0.3,sloped,above]{0.72} (v3);
        \draw[edge_style]  (v1) edge node[pos=0.47,sloped,above]{0.65} (v4);
        \draw[edge_style]  (v1) edge node[sloped,above]{0.74} (v5);
        \draw[edge_style]  (v1) edge node[above,pos=0.6,sloped,,above]{0.31} (v6);
        \path (v2) edge [loop above,font=\footnotesize] node {15.52} (v2);
        \draw[edge_style]  (v2) edge node[sloped,above]{1.04} (v3);
        \draw[edge_style]  (v2) edge node[auto,swap,pos=0.3,sloped,above]{0.81} (v4);
        \draw[edge_style]  (v2) edge node[auto,swap,pos=0.4,sloped,above]{0.80} (v5);
        \draw[edge_style]  (v2) edge node[swap,pos=0.51,sloped,above]{0.34} (v6);
        \path (v3) edge [loop above,font=\footnotesize] node {8.05} (v3);
        \draw[edge_style]  (v3) edge node[pos=0.4,sloped,above]{0.40} (v4);
        \draw[edge_style]  (v3) edge node[pos=0.62,sloped,above]{0.21} (v5);
        \draw[edge_style]  (v3) edge node[sloped,above]{0.13} (v6);
        \path (v4) edge [loop below,font=\footnotesize] node {11.09} (v4);
        \draw[edge_style]  (v4) edge node[sloped,above]{0.60} (v5);
        \draw[edge_style]  (v4) edge node[sloped,above]{0.42} (v6);
        \path (v5) edge [loop below,font=\footnotesize] node {19.88} (v5);
        \draw[edge_style]  (v5) edge node[pos=0.53,sloped,above]{0.39} (v6);
        \path (v6) edge [loop right,font=\footnotesize] node {11.33} (v6);
    \end{tikzpicture}
     \caption{Fraction of intra-shard and cross-shard transactions in obtained partition with six worker shards $X_1$ to $X_6$. Self-edge weights denote the fraction of intra-shard transactions (out of all transactions), while other edge weights denote the fraction of transactions (out of all transactions) involving those two shards.}
     \label{fig:tx-dist}
\end{figure}

\two 
\noindent
{\bf Intra and cross-shard activities.}
We observe the fraction of cross-shard transactions as well as intra-shard transactions to confirm that the partitioning indeed leads to a reduction of cross-shard activity. 
Figure~\ref{fig:tx-dist} illustrates the results.  
The number on an edge is the fraction of transactions involving those two shards (possibly others).
The number on a self-edge is the fraction of transactions involving only that shard.
As anticipated, there are much fewer  
cross-shard transactions than intra-shard transactions.

On a deeper look within each shard, we observe that each shard has a few popular contracts that lead to high
intra-shard activities. 
For example, all the accounts of a popular 
cryptocurrency exchange Binance~\cite{binance}, are assigned to shard $X_1$, and 
these accounts frequently interact with each other. Shard $X_2$
has Ethermine~\cite{ethermine}, a popular Ethereum mining pool, that frequently pays
the miners in the pool whose accounts are mostly assigned to 
the same shard. Shard $X_3$'s most popular account is the ``Tether token''~\cite{tether} contract, a popular ERC20 token with a value 
pegged to the US dollar.

\begin{figure}[bt]
    \centering
    \pgfplotsset{footnotesize,height=4.5cm, width=0.85\linewidth}
    \begin{tikzpicture}
    \begin{axis}[
        legend pos=south east,
        legend columns=2,
        xtick=\empty,
        extra x ticks={32,64,96,128,160,192,224,256,288},
        xlabel={Required data (in bytes)},
        ylabel= {Fraction of transactions},
        ] 
		\addplot [mark=x,orange] table [x=keysize, y=s1, col sep=comma] {data/cdataFreq.csv};
		\addplot [mark=+,cyan] table [x=keysize, y=s2, col sep=comma] {data/cdataFreq.csv};
		\addplot [mark=o,brown] table [x=keysize, y=s3, col sep=comma] {data/cdataFreq.csv};
		\addplot table [x=keysize, y=s4, col sep=comma] {data/cdataFreq.csv};
		\addplot table [x=keysize, y=s5, col sep=comma] {data/cdataFreq.csv};
		\addplot table [x=keysize, y=s6, col sep=comma] {data/cdataFreq.csv};
        \addlegendentry{Shard $X_1$}
        \addlegendentry{Shard $X_2$}
        \addlegendentry{Shard $X_3$}
        \addlegendentry{Shard $X_4$}
        \addlegendentry{Shard $X_5$}
        \addlegendentry{Shard $X_6$}
    \end{axis}
    \end{tikzpicture}
    \caption{Cumulative fraction of cross-shard transactions in all 
    the six shards in terms of data downloaded from other shards (in bytes).}
    \label{fig:key-freq}
\end{figure}
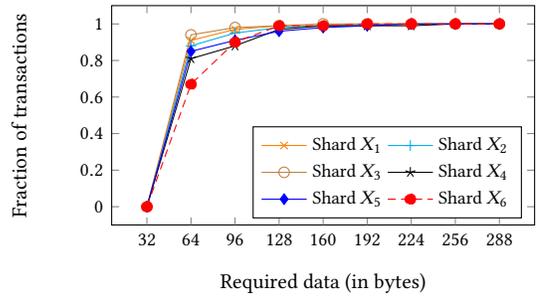

\two
\noindent
{\bf Data transfer between shards.} 
Since every \ws\ shard maintain disjoint subset of the entire
state, sometimes nodes of a \ws\ shard need to download state
information from nodes
of other \ws\ shards to execute cross-shard transactions.
Figure~\ref{fig:key-freq} illustrates the cumulative 
distribution of cross-shard transactions in terms of the 
amount of data transfer needed. 
Observe that, in every \ws\ shard, 
more than 95\% of cross shard transactions only require transferring at most 128 bytes of data (4 values) from other shards 
to execute the transaction locally. This shows that the data transfers for local execution of cross shard transactions is minimal.
%

\two
\noindent
{\bf Potential gaps from future sharded blockchains.}
Despite our efforts to mimic realistic workloads, we would like to acknowledge the potential gap between our workload and real-world sharded blockchains (when they come into existence).
Since cross-shard transactions are inherently more expensive (involve locking and data transfer between shards), users may take intelligent measures to reduce the amount of cross-shard transactions they use. 
For example, a user who repeatedly uses a service from a shard other than his home shard may decide to create accounts in that other shard and transfer some tokens to it. 
Further, applications or contracts that expect to receive a large number of cross-shard transactions 
might adopt a programming practice to distribute its address space to reduce conflicts between different transactions. These behaviors and practices may lead to a further reduction in cross-shard transactions in comparison to our benchmark.
The dominating activities on the Ethereum blockchain (and other blockchains) today come from trading, exchanges, and mining pools. This will likely change if blockchains are to find more practical applications.
It is hard to predict what applications will prevail and what characteristics (related to sharding) they will exhibit. 
The methodology in this section represents our best effort in creating a sharded blockchain workload given the data available at the time of writing.

\section{\prot\ Design}
\label{sec:design}
\begin{figure*}[t!]
    \centering
    \includegraphics[width=\linewidth]{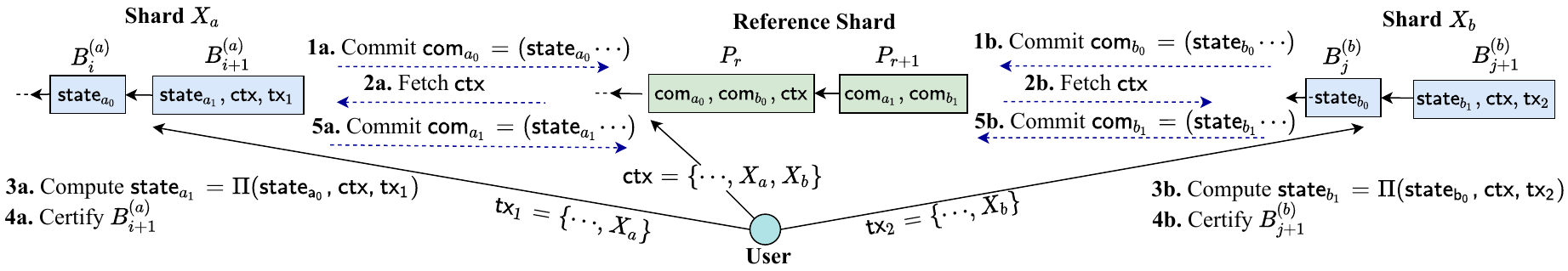}
    \caption{Overview of \prot\ with two shards $X_a,X_b$ 
    and a cross-shard transaction $\ctx$ involving both shards. 
    $\tx_1$ and $\tx_2$ are intra-shard transactions to $X_a$
    and $X_b$ respectively. Here $B^{(a)}_i$ and $B^{(a)}_{i+1}$~(resp. $B^{(b)}_j$ and $B^{(b)}_{j+1}$)
    are the two certified \ws\ blocks at shard $X_a$ (resp. $X_b$). Actions taken by $X_a$ (resp. $X_b$) are indicated as {\bf 1a}, $\cdots$ , {\bf 5a} (resp. {\bf 1b}, $\cdots$, {\bf 5b}) and they occur in the specified sequence. Here we overload notation to 
    use $\st_{a}$ as the cryptographic digest of $\st_{a}$.
    Note that commitment $\cm$ of worker shard blocks includes other information in addition to the corresponding state. We represent them with ellipses~($\cdots$) in this figure as they are not relevant for understanding the overview.}
    \label{fig:overview}
\end{figure*}

\subsection{System Model}
\label{sub:network}
We adopt the standard partially synchronous network model, i.e., a network that oscillates between periods 
of synchrony and periods of asynchrony. During periods of synchrony 
all messages sent by honest replicas adhere to to a known delay bound 
$\Delta$. During periods of asynchrony messages, messages can be delayed arbitrarily. In theoretical works, the partial synchrony model~\cite{dwork1988consensus} is often stated differently (e.g., using an unknown Global Standardization Time, GST) for rigor or convenience, but the essence is to capture the practical oscillating timing model mentioned above.
A protocol in partial synchrony ensures safety (consistency) even under periods of asynchrony, and provides liveness only during periods of synchrony.

This paper focuses 
on the permissioned setting.\footnote{It is conceivable to have (some or all) shards to be permissionless by running a permissionless blockchain per shard. We leave this direction as future work.}
We assume that at most $f$ replicas can be faulty in each shard. 
A consensus protocol tolerating $f$ faults under partial synchrony needs at least $3f+1$ replicas~\cite{dwork1988consensus}.
Thus, in \tpc-based protocols, every shard has $3f+1$ replicas.
In \prot, only the \rs\ shard has $3f+1$ replicas and every \ws\ shard has $2f+1$ replicas.
All faulty replicas are controlled 
by a single adversary \adv\ and they can deviate arbitrarily 
from the prescribed protocol. 
All non-faulty replicas are honest 
and they strictly follow the prescribed protocol. 
We assume that \adv\ cannot break standard cryptographic constructions such as hash functions and signatures schemes.


\subsection{Overview}
In \prot\, there are $k+1$ shards $\{X_0,X_1,\cdots,X_k\}$ in total.
Shard $X_0$ is referred to as the {\em \rs} 
shard and runs a fault tolerant consensus protocol. 
All cross-shard transactions in \prot\ are included and ordered by the \rs\ shard. 
The other $k$ shards are called
{\em \ws} shards, and they maintain disjoint subsets of the system 
states.
The \ws\ shards verify and prove the validity of its state;
but they do not need to provide consensus.
Hence, as mentioned, the \rs\ shard has $3f+1$ replicas while
each \ws\ shard has $2f+1$ replicas.

Each \ws\ shard maintains a sequence of certified blocks where 
each block includes some intra-shard transactions and a hash of 
its predecessor. We refer to 
this sequence of blocks as the \ws\ chain. Replicas within 
a \ws\ shard append certified blocks to the \ws\ chain by 
collecting at least $f+1$ distinct signatures from replicas
within the shard. Once a block is certified, the \ws\ shard
submits a commitment of the resulting state to the \rs\ chain, to be finalized by the \rs\ shard. We note again that, instead  
of running a consensus protocol per \ws\ shard, \prot\ 
only requires \ws\ shards to certify the validity of \ws\ blocks per the 
protocol specification~(see~\S\ref{sub:ws-block} for precise 
definition of valid blocks). 
A \ws\ block is finalized when a \rs\ 
block containing its commitment is finalized in the \rs\ chain. 




Figure~\ref{fig:overview} 
illustrates the high-level idea behind \prot\ with an example.
Say a user creates a cross-shard transaction $\ctx$ that involves two shard $X_a$ and $X_b$. 
Let $\st_{a_0}$ and $\st_{b_0}$ be the latest committed states
from $X_a$ and $X_b$ respectively. 
Also, let $\tx_1$ and $\tx_2$ be two intra-shard transactions. 
%
Here, $\ctx$ is first included in block $P_r$  by the \rs\ shard; then replicas 
in $X_a$ and $X_b$ execute $\ctx$ atop the latest committed states $\st_{a_0}$ and $\st_{b_0}$. After executing $\ctx$, both shards independently execute some intra-shard transactions, 
e.g., $X_a$ executes $\tx_1$ and $X_b$ executes $\tx_2$, and update their 
commitments of the latest execution results $\st_{a_1}$ and $\st_{b_1}$. 

A careful reader may note that the core approach in \prot\ can
be viewed as a locking scheme. At every state
commitment, each shard implicitly locks its entire state to potential
future cross-shard transactions. 
However, despite locking the 
state, a \ws\ shard optimistically proceeds to execute and certify new 
intra-shard transactions atop the locked state, hoping that no conflicting cross-shard 
transactions will appear in the \rs\ chain before they commit 
the updated state.
If indeed no cross-shard transactions involving a \ws\ shard 
appear in the \rs\ chain, the new state commitment gets added 
to the \rs\ chain and the \ws\ shard makes progress. 
On the other hand, if some conflicting cross-shard transactions appear before 
the next state commitment, \prot\ forces \ws\ shards to 
execute those cross-shard transactions first before any new intra-shard
transactions. In doing so, a \ws\ shard may have to discard some certified blocks in its \ws\ chain. We report statistics on how 
often a \ws\ shard has to discard its certified blocks in~\S\ref{sec:eval}.


Note that every \ws\ shard can independently 
execute \emph{cross-shard} transactions as soon as it notices them in a 
finalized \rs\ block, independent of the status quo of other involved shards.
This is in sharp contrast to \tpc\, where each shard waits for 
every other shard specified in the transaction to lock its 
state first, and only then proceeds to execute the cross-shard 
transaction atop the locked states. Indeed, this very nature 
of pro-active state commitments allows \prot\ to execute cross-shard 
transactions more efficiently than \tpc.

\subsection{Data structures}
\label{sub:struct}
\noindent
{\bf \Ws\ shard blocks.}
A certified \ws\ block $B_i$ at height $i$ at shard $X_a$ consists of the 
following components
\begin{equation}
    B_i = \la \hs_{i-1}, r_i, \st_i, \bt_i \ra
\end{equation}
Here, $\hs_{i-1}$ is the hash of the parent \ws\ block, $r_i$ is 
the height of the latest known \rs\ block, and $\bt_i$ is a ordered 
list of intra-shard transactions. Let $\bq^{(a)}$ be 
the ordered list of cross-shard transactions that are included 
in the \rs\ block since the last commitment from $X_a$ to 
height $r_i$ (both inclusive) and involve $X_a$. Then, $\st_i$, 
the state at the end of $B_i$, is the resulting state after 
executing $\bq^{(a)}$ followed by transactions in $\bt_i$. 
When $B_{i-1}$ with state $\st_{i-1}$ is the latest 
committed block from $X_a$:
\[
    \st_i = \Pi\left(\Pi\left(\st_{i-1}, \bq^{(a)}_{r_i}\right), \bt_i\right)
\]
The {\em certificate} of a \ws\ block is a signature from at
least $f+1$ distinct replicas within the shard. 


%
%

\two 
\noindent
{\bf \Rs\ shard blocks.}
A \rs\ shard block $P_r$ at height $r$ consists of 
the following components:
\begin{equation}
    P_{r} = \la \hs_{r-1},\bc_r,\bq_r \ra
\end{equation}
Similar to \ws\ blocks, $\hs_{r-1}$ is the hash of the parent \rs\ block, 
$\bc_r$ is a list of block commitments from (not necessarily all) \ws\ 
shards, $\bq_r$ is an ordered list of new cross-shard transactions.
Sometimes, we use $\bq_{r,s}$ to indicate the ordered list of 
cross-shard transactions that appear between \rs\ blocks at 
height $r$ and $s$ (both inclusive). 

We explain the contents of the block commitments along with their 
purpose and description of cross-shard transactions in the upcoming paragraphs. 

\two 
\noindent
{\bf Block commitments.}
\Ws\ shards generate a block commitment denoted as $\cm$ after 
every \ws\ block and broadcast it to the \rs\ shard. 
Block commitments in \prot\ serve two purposes. First, 
\rs\ shard uses these commitments to order blocks inside \ws\ shard; 
second, commitments from a shard also acts a promise to every other shard that 
all future cross-shard transactions will be executed atop the latest committed
state. 
Also, as some commitments can get delayed due to network delay, 
\prot\ allows \ws\ shards to certify newer blocks and directly 
submit commitment of any successor block of the latest committed
block. Specifically, for a block $B_i$ at height 
$i$, its commitment $\cm_i$ consists of a cryptographic digest of $\st_i$, the resulting state after 
$B_i$ and sequence of a hash chain $\bh_i$ of certified block hashes 
starting with the hash of the last committed block to the current block
$B_i$, i.e., $\cm_i=\la \st_i,\bh_i \ra$.
Replicas in the \rs\ shard use this hash chain 
to validate that the block indeed extends the last finalized block from that \ws\ shard. 
When clear from the context, we overload the 
notation $\st_i$ to denote the cryptographic
digest of $\st_i$.

\two 
\noindent
{\bf Intra-shard transaction.} 
Intra-shard transactions in \prot\ specify the identity of the function they wish to invoke and appropriate 
function parameters. Creators of intra-shard transactions send these 
transactions to replicas of the \ws\ shard storing 
the state required for their execution. Respective \ws\ shard replicas 
then gossip the transactions among themselves and include it in the 
next available \ws\ block.

\two 
\noindent
{\bf Cross-shard transaction.}
In addition to information specified in every intra-shard transactions, 
every cross-shard transaction also specifies its potential read-write
set in its description. 
The creator of every cross-shard transactions use ideas akin to 
Optimistic Lock Location Prediction~(OLLP)~\cite{thomson2012calvin} to
generate the read-write set. We include the read-write set in the 
description of the cross-shard transaction to indicate the subset of 
shards necessary for executing the transaction along with the keys these 
shards need to exchange for its execution. Also, unlike intra-shard 
transactions, creators of every cross-shard transaction send their 
transaction directly to at least one honest replica of the \rs\ shard. 
These replicas then gossip these transactions among themselves and 
include them in new \rs\ blocks as described in the next section. 

%


%
\subsection{\Rs\ Shard Protocol}
\label{sub:rs-shard} 
Replicas in the \rs\ shard run a standard consensus protocol,
such as PBFT~\cite{castro1999practical} or HotStuff~\cite{yin2019hotstuff}, to finalize new proposed blocks and append them to the \rs\ chain.
For concreteness, we use Hotstuff~\cite{yin2019hotstuff} as the 
underlying consensus protocol in the \rs\ shard. In this section, 
we primarily focus on the rules for proposing a new block, as we
use the remaining part of the Hotstuff protocol as it is. 

As in HotStuff, we use {\em views} with one leader per view. 
In every view, the leader of that view is responsible for 
driving consensus on newer blocks. 
Let $L$ be the leader of the current view. To propose a new \rs\ 
block $P_r$ at height $r$, $L$ includes a subset of valid block 
commitments and cross-shard transactions. In $P_r$, the state commitments 
of \ws\ shards are ordered before the cross-shard transactions, 
and they are chosen as follows. 

Let $X$ be a shard with $\cm_l$ with state $\st_l$ for block $B_l$ 
as its latest commit that appears in \rs\ chain up to the parent 
block of the proposal $P_r$. 
Let $\cm_l$ appear in \rs\ block $P_s$. 
Then, a new of commitment $\cm_j= \la \st_j,\bh_j \ra$ for a block 
$B_j$ reporting a \rs\ block at height $s$ from $X$ is valid if 
and only if: 
\begin{enumerate}
    \item Each \ws\ shard block whose hash appear in the hash chain $\bh_j$ has been signed by at least $f+1$ distinct replicas in $X$; and 
    \item The block $B_j$ extends the latest committed block $B_l$ of the
    shard; $L$ validates this using the hash chain $\bh_j$ mentioned inside 
    the commitment $\cm_j$.
    \item No cross-shard transaction involving $X$ appears after \rs\ block $P_s$ up until the parent block of $P_r$. 
\end{enumerate}

%

\begin{figure}[t!]
    \centering
    \includegraphics[width=0.90\linewidth]{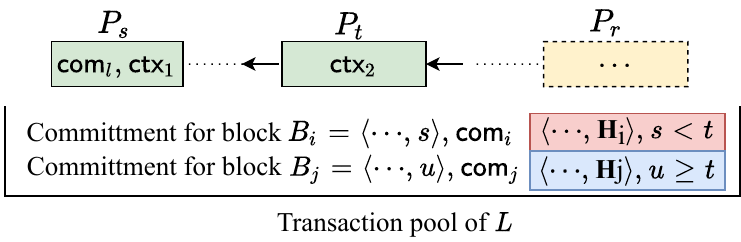}
    \caption{Illustration of valid~($\cm_j$) and invalid~($\cm_i$) 
    block commitments from \ws\ shard $X$ available at a \rs\ 
    chain block proposer $L$ prior to its new proposal $P_r$. 
    Here, $\cm_l$ is the latest known block commitment 
    from $X$ that had been included in \rs\ block $P_s$.
    $P_t$ is the latest \rs\ block that includes the cross-shard
    transaction $\ctx_2$ involving $X$.}
    \label{fig:commit-valid}
\end{figure}

For example, in Figure~\ref{fig:commit-valid}, $P_r$ at height $r$ 
is the block $L$ wants to propose and $P_s$ at height $s$ is 
the \rs\ block that includes the latest commitment $\cm_l$ of 
shard $X$. Also, let $P_t$ at height $q$ be the last \rs\ block that includes a cross-shard transaction involving $X$. 
Let $\cm_i$ and $\cm_j$ be 
two newly available commitments from $X$ then commitment $\cm_i$ 
is invalid as its violates the third condition mentioned above. 
Specifically, $s$ reported in $\cm_i$ is less than $t$, and 
$P_t$ includes $\ctx_2$ a cross-shard transactions involving 
$X$. On the other hand, assuming $\bh_j$ is a valid hash chain, $\cm_j$ 
is a valid state commitment since $u\ge t$. We summarize the 
procedure for validating state commitments in 
Algorithm~\ref{algo:ref-propose}.

For every cross-shard transaction $\ctx$ in $L$'s transaction pool, 
it is considered valid if and only if $\ctx$ does not intend to 
read and write from a key that some preceding cross-shard transaction already
intends to write. 
Let $\bx_{\ctx}=\{X_1,X_2,\cdots,X_m\}$ be the subset of shards that
are involved in executing $\ctx$. Let $\br^{(a)}_{\ctx}$ be the set of 
keys from shard $X_a \in \bx_{\ctx}$ that $\ctx$ mentions in 
its read-set. Also, let $\cm_{a}$ be the latest block commitment
by shard $X_a$. Let $\bw^{(a)}$ be the set of 
keys mentioned in the write set of cross-shard transaction
that are included after the commit $\cm_{a}$. Then, $\ctx$ 
is considered valid if and only if, 
\begin{equation}
    \br^{(a)}_{\ctx} \cap \bw^{(a)} = \emptyset,\ \ \forall X_a \in \bx_{\ctx}
\end{equation}

Stated differently, this validation check ensures that every
cross-shard transaction reads keys that have not been written
to by any other cross-shard transaction since the last commit. 
Refer to Algorithm~\ref{algo:ref-propose} for precise details.
This is important as it enables replicas in a \ws\ shard to prove
and validate the correctness of data they exchange during execution 
of cross-shard transactions~(see~\S\ref{sub:execute cst}).

\begin{figure}[t!]
    \centering
    \includegraphics[width=\linewidth]{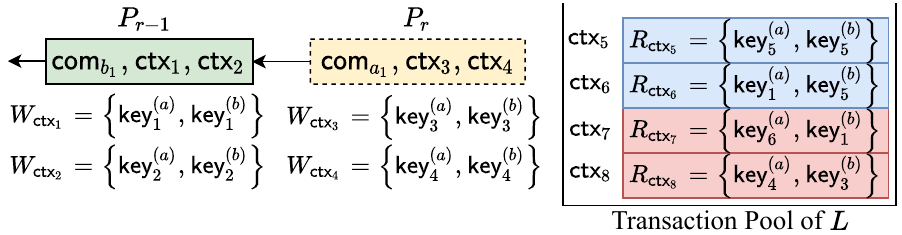}
    \caption{Illustration of valid~($\ctx_5,\ctx_6$) and 
    invalid cross-shard transactions in the transaction
    pool of the current \rs\ chain leader $L$. 
    Validation of cross-shard transactions by leader $L$ before including them in the next \rs\ block $P_{r}$. 
    Transactions shaded in blue are the set of valid transactions 
    and transactions shaded in red are the set of invalid 
    transactions.}
    \label{fig:cross-valid}
\end{figure}

Figure~\ref{fig:cross-valid} provide an illustration of how 
the proposer $L$ validates the cross-shard transactions it
observes to include them in its next proposal. 
Let $P_{r-1}$ be the latest \rs\ block with cross-shard
transactions $\ctx_1$, $\ctx_2$ known to $L$. 
Also, let $\cm_{a_1}$ and $\cm_{b_1}$ be the latest commitments from 
shards $X_a$ and $X_b$ respectively. Say $L$ wants to propose the 
next block $P_{r}$. Let us assume that keys
$\key^{(a)}_i$'s and $\key^{(b)}_i$'s 
denote the states maintained by shard $X_a$ and $X_b$
respectively. Let
$R_{\ctx}$ and $W_{\ctx}$ denote the read-write set
mentioned in the description of the cross-shard transaction
$\ctx$. Lastly, let's assume that $L$ has already included $\cm_{a_1},\ctx_3$, and $\ctx_4$ in the $P_r$ it has created so far.

Now, among the remaining transactions from transaction pool
of $L$, i.e., $\{\ctx_5,\ctx_6,\ctx_7,\ctx_8\}$ in our example, 
$\ctx_7$ and $\ctx_8$ can not be included in $P_r$ as $\ctx_7$ 
aims to read from  $\key^{(b)}_1$ on which $\ctx_1$ already 
holds a write-lock. Similarly, $\ctx_8$ aims to read from keys 
$\key^{(b)}_3$ and $\key^{(b)}_4$ that are in write 
set of $\ctx_3$ and $\ctx_4$ respectively. On the contrary, 
$\ctx_5$ and $\ctx_6$ do not have any read-write conflicts 
with any of the cross-shard transactions included so far.

\subsection{\Ws\ Shard Protocol}
\label{sub:ws-block}
\begin{figure*}
\small
\begin{mdframed}
In every view $v$,
    \begin{enumerate}
        \item {\bf Propose.} Leader of view $v$, $L$ creates a new 
        block $B_i$ at height $i$ following Algorithm~\ref{algo:ws-propose} and broadcasts it to the all replicas within the shard.
        
        \item {\bf Certification.} On hearing the proposal $B_i$, non-leader
        replicas validates $B_i$ by running Algorithm~\ref{algo:ws-valid}. 
        On successful validation, the replica signs $B_i$ and send 
        the signature to $L$. $f+1$ distinct valid signatures on $B_i$
        is called a certificate of $B_i$. Once $B_i$ is certified, $L$
        creates its commitment $\cm_i$ and broadcasts it to the reference
        shard.
    \end{enumerate}
\end{mdframed}
\caption{Summary of worker shard protocol in a view $v$.}
\label{prot:worker}
\end{figure*}

Although the protocol for a \ws\ shard is not a consensus protocol, 
we will borrow ideas from popular leader-based paradigm in 
consensus protocols. In particular, similar to \rs\ shard protocol, 
the \ws\ shard protocol proceeds in views. Views are numbered by
monotonically increasing integers. Also, for each view $v$, one 
\ws\ replica say replica with identity $v\% n$ serves as the 
leader of the view. The leader is responsible for proposing new
\ws\ blocks, getting them certified by the \ws\ shard, and 
submitting them to be finalized in the \rs\ chain. 

Similar to Hotstuff~\cite{yin2019hotstuff}, we use the rotating
leader approach for \ws\ shard, i.e., views are incremented 
after every fixed interval and the appropriate node is chosen 
as the leader of the new view. By doing so, we obviate the 
need of explicit leader-replacement protocol that are required
for protocol such as PBFT~\cite{castro1999practical}. 

Next, we describe the detailed protocol within a view in
Figure~\ref{prot:worker}. Within each view the protocol 
has two phases: a \emph{proposal} phase and a 
{\em certification} phase.

\two 
\noindent
{\bf The proposal phase.} 
Leader $L$ of current view in a \ws\ shard $X$ proposes a new 
block $B_i$ at height $i$ by broadcasting a {\em propose} message 
to other replicas within the shard. 
Recall, each new proposal reports a state $\st$ after 
executing all the cross-shard transactions~(if any) known 
to the leader, followed by some intra-shard transactions. 
All these transactions are executed atop the last 
committed state of the shard. 

When latest known \ws\ block $B_{i-1}$ of shard 
$X$ is already committed, $L$ extends it by first executing 
cross-shard transactions that appear since commitment of $B_{i-1}$ 
atop the committed state and then it executes some intra-shard 
transactions. Alternatively, when the latest known block 
$B_{i-1}$ is not yet committed, $L$ extends $B_{i-1}$ if and only 
if no cross shard transactions appear since the \rs\ block, $P_{r_{i-1}}$, reported in $B_{i-1}$. Otherwise, $L$ proposes 
a new block atop the latest committed block, say $B_j$, from 
shard $X$, after executing all the cross-shard transactions 
known since the last commitment.

\begin{figure}[tb]
    \centering
    \includegraphics[width=\linewidth]{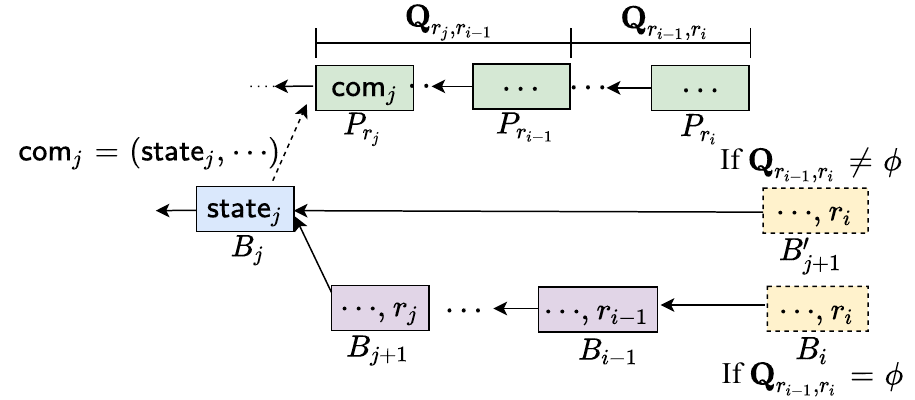}
    \caption{Illustration of the protocol followed by a honest proposer 
    $L$ in \ws\ shard to propose a new block. Here $B_j$ with state $\st_j$ 
    is the latest committed block. Let the commitment for block
    $B_j$, $\cm_j$, is included in the \rs\ block $P_{r_j}$ and
    $B_{r_i}$ be the latest \ws\ block known to $L$.}
    \label{fig:reorg}
\end{figure} 
Figure~\ref{fig:reorg} illustrates this through an example
where $L$ proposes the new block atop $B_{i-1}$ only if 
$\bq_{r_{i-1},r_i}$ is empty, i.e., no new cross-shard 
transactions involving $X$ appear after \rs\ block $P_{r_{i-1}}$.
Otherwise, $L$ proposes the next \ws\ block atop the latest
committed block $B_j$ from worker shard, after executing 
all transactions in $\bq_{r_j,r_i}$. Recall, $\bq_{r_j,r_i}$ 
denotes the set of relevant cross-shard transactions that 
are included in a reference block since the inclusion of 
the $\cm_j$ in the reference block $P_{r_j}$. We summarize this
in Algorithm~\ref{algo:ws-propose}.

\two
\noindent
{\bf The certification phase.} 
Each honest replica $n$ upon receiving the proposal $B_i$, replies 
with a {\em vote} message if the replica is in the same view as the 
proposal and the proposal is valid. A valid proposal satisfies 
the following properties, which we summarize in Algorithm~\ref{algo:ws-valid}.
\begin{enumerate}
    \item $B_i$ extends $B_l$, the latest committed block known to $n$ 
    and the \rs\ block known to $n$ is at a height greater than or 
    equal to the \rs\ block mentioned in $B_i$. 
    \item The state mentioned in the proposal satisfies the properties
    of the honest proposal mentioned earlier. 
\end{enumerate}


On receiving $f+1$ distinct valid signatures from $f+1$ 
distinct replicas, the leader $L$ aggregates them 
into a certificate, and sends the block commitment $\cm_i$ for 
$B_i$ to the \rs\ shard. It is then the responsibility of the
replica's in the \rs\ shard to include $\cm_i$ in the next 
available \rs\ block. As mentioned earlier, once the commitment 
$\cm_i$ or a commitment of its successor block is included in 
the \rs\ chain, the \ws\ block $B_i$ is finalized.

Informally, since there are at most $f$ Byzantine replicas
in each \ws\ shard, $f+1$ distinct signatures on $B_i$ implies
that at least one honest replica has validated the $B_i$
as per Algorithm~\ref{algo:ws-valid} and checked that $B_i$
satisfies all the requirement. This ensures that every 
certified block adheres to the protocol specification. 
Similarly, since there exists at least $f+1$ honest replicas
in each every worker shard, the requirement of only $f+1$ 
signatures ensures that honest leaders can successfully 
create the required certificate with the help of only 
honest replica. We will formally argue this later in. 

All worker shard replica then enter the next view after 
a pre-specified time interval and the cycle continues.

\subsection{Execution of Cross-shard Transaction}
\label{sub:execute cst}
Once a cross-shard transaction $q$ appears in the \rs\ chain, the
shards involved in its execution,  $\bx_q=\{X_1,\cdots,X_m\}$, 
exchange the values corresponding to keys mentioned in the read set 
of $q$ with each other. Specifically, replicas within every shard
$X\in \bx_q$ request replicas of remaining shard for the committed
values of addresses mentioned in the read-set of $q$ that are not
maintained by $X$ . On receiving responses from shards in 
$\bx_q$, each replica validates the received value against the
appropriate state commit. Upon correct validation, the proposer 
of the next \ws\ block executes the cross-shard transactions in 
the order they appear in the \rs\ block. 

To avoid data download on every cross-shard transaction, \prot\ 
batches cross-shard transactions and sends a single download request for all the keys used 
in all transactions in one \rs\ block. Also, during
execution, each shard updates its local state whenever
the transaction writes to keys from the local shard.

A few subtleties arise in this process. 
First, every shard should be aware of the description of 
every function that gets executed as a part of each cross-shard
transaction. For example, if a cross-shard transaction executes
functions from two different shards, both shards should be aware 
of the function description. \prot\ addresses this by tagging each 
smart contracts as {\em global} or {\em local}. Every shard 
stores descriptions of all global contracts. 
\prot\ uses cross-shard transactions to create global contracts
and cross-shard transactions in \prot\ can only invoke functions 
of global contracts. Local smart contracts are created using 
intra-shard transactions and they only accept intra-shard 
transactions. 
Although this may appear to result in considerable overhead, this can be 
avoided by better programming practices. As illustrated in~\cite{kiffer2018analyzing,he2019characterizing}, most 
of the contracts in Ethereum are copies of each other. Hence, a 
better programming practice would be to create standard global 
libraries for common functionalities such as  ERC'20 and Exchanges.

The second subtlety arises from potential mismatch in the read-write 
set mentioned in the description of a transaction and the 
read-write set accessed by the transaction during its 
execution within the \ws\ shards. In such scenarios, each 
replica aborts execution of the transaction, reverts all the changes 
caused by its execution so far, and proceeds to the next cross-shard transaction.

\section{Analysis}
\label{sec:analysis}
In this section, we provide a proof sketch of the safety and liveness guarantees of \prot. 
The detailed proofs are deferred to Appendix~\ref{apx:security}.
We will then analyze the performance of \prot\ and compare it
against the \tpc\ based approach.

\subsection{Safety and Proof Sketch of Liveness}
\label{sub:ws correctness}
Safety of \prot\ follows directly from the 
safety of the Byzantine fault tolerant consensus protocol used in 
the \rs\ shard. This holds true even during periods of asynchrony
as a partially synchronous consensus algorithm is safe under asynchrony. 
To elaborate, the consensus algorithm in the \rs\ shard provides global order for all transactions, the intra-shard ones as well as cross-shard ones.
Each transaction is associated with a unique \rs\ block that finalizes it in the \rs\ chain.
Transactions are hence ordered first by their heights in the \rs\ chain, and then by their positions inside the \rs\ block. 

Besides an agreed upon total order, \prot\ also ensures every \ws\ block commitment finalized in the \rs\ chain represents a valid state. 
The \rs\ shard ensures that at 
most one \ws\ block at any given height from a \ws\ shard gets finalized and it extends a previously finalized \ws\ block from that shard. 

We next show that \prot\ makes progress during periods of synchrony,
i.e., when messages between honest replicas get delivered
within a bounded delay of $\Delta$. 
Specifically, during periods of synchrony, the \rs\ chain and each \ws\ chain will make progress. 
To see this, consider the shard $X$ and let $P_r$ be the latest 
\rs\ block, and $L_r$ be honest leader for the next view. 
Let $t$ be the time instant when $P_r$ is created. This implies 
that by time $t + \Delta$, $L_r$ will know about $P_r$.
Hence, every honest replica of shard $X$ including the proposer of 
the next block, $L$, will be aware of the block $P_r$ by time 
$t+2\Delta$. 
Also, since each replica of every shard is connected with at least 
one honest replica of every other shard, by time $t+4\Delta$ every 
honest replica of shard $X$ will have the required state to execute
the cross-shard transactions in blocks up to $P_r$. 

Hence, when $L$ proposes the next block at time $t+4\Delta$,
every honest replica will respond immediately with its signature.
Thus, by time $t+6\Delta$, $L$ will collect a certificate for 
its proposal, and by time $t+7\Delta$, the block commitment will 
reach an honest replica of \rs\ chain. Also, by time $t+8\Delta$ 
it will reach the leader of the \rs\ chain. This implies that the 
commitment of \ws\ block or one of its successors will appear in 
the next \rs\ block. Refer to Appendix~\ref{apx:security} 
for detailed proof.

\subsection{Performance Analysis}
\label{sub:comm}
It is easy to see that in both \prot\ and \tpc\ based approach,
each \ws\ shard only stores a subset of the entire state. Hence, 
both approaches achieve state sharding. Also, each \ws\ shard 
validates a subset of all intra-shard transactions and the 
cross-shard transactions it is a part of. Hence, both protocols 
achieve computation sharding. Moreover, since the \rs\ 
shard runs a standard BFT conensus protocol, the communication 
complexity of finalizing a \rs\ block is same as the communication 
complexity of the underlying consensus protocol, e.g., HotSuff
only requires linear communication. 

Contrary to \rs\ blocks, finalization of a \ws\ block $B_i$ in \prot\
involves two steps: certification of $B_i$ and finalization of 
the \rs\ block that includes the commitment $\cm_i$ of $B_i$.
It is easy to see from~\S\ref{sub:ws-block} that certification of 
every \ws\ block involves only one round of communication: the 
leader broadcasts a new proposal to each replica and they 
respond with their signatures. Since both  these steps have
linear communication costs, overall block certification protocol 
has linear costs as well. Hence, assuming a linear consensus protocol
in the \rs\ chain, the overall communication of finalizing a \ws\ 
block is also linear. An important point to note is that 
each \rs\ block will potentially include numerous block commitments
and cross-shard transactions simultaneously, and hence the 
communication overhead gets amortized. 

\two 
\noindent
{\bf Confirmation latency of transactions.}
The cross-shard transaction $\ctx$ is finalized as soon as
$\ctx$ gets included in a \rs\ block. 
The state 
atop which $\ctx$ should be executed has also been finalized by then.
The only thing remaining 
is to get the actual execution result, i.e., the resulting state modification due to executing $\ctx$. 
Since every shard involved executes $\ctx$ deterministically
atop an identical state, this execution result becomes
available as soon as one \ws\ shard block 
containing $\ctx$ gets certified. 
It does not matter which participating \ws\ worker shard first does so.
Hence, the \emph{confirmation latency} 
of a cross-shard transaction is measured as the 
time elapsed since its inclusion in the \rs\ chain till the first \ws\ block containing it gets certified.

We measure the confirmation latency of an intra-shard transaction as the elapsed time between its inclusion in a  
\ws\ block and the finalization of  \ws\ shard block. 
Note that in \tpc, \ws\ blocks are finalized immediately and so are the 
intra-shard transactions in them.

\section{Implementation \& Evaluation}
\label{sec:eval}
We implement \prot\ and \tpc\ atop the open-source 
\qrm\ client version ${\sf 2.4.0}$~\cite{jpmorgan2020quorum}. \qrm\ is a 
fork of the Ethereum Go client and inherits Ethereum's smart contract execution platform and implements a permissioned consensus protocol based on the Istanbul BFT~(IBFT) and Tendermint~\cite{buchman2016tendermint}
consensus algorithm.\footnote{Saltini and Hyland-Wood in~\cite{saltini2019correctness} discusses a liveness bug in the
original design of IBFT. The bug has been fixed since then in~\cite{saltini2019ibft}.}
For \prot, we use the \ibft\ 
implementation for the \rs\ shard, and we implement the protocol described 
in~\S\ref{sub:ws-block} for each \ws\ shard.
For \tpc, we use the \ibft\ 
implementation for all shards. 

Given the exploding number of sharding proposals~\cite{danezis2015centrally,
kokoris2018omniledger,al2017chainspace,zamani2018rapidchain,wang2019monoxide,dang2019towards,manuskin2019ostraka}, 
it is difficult to replicate each of their unique (and vastly different) parameter settings and system model.
Since they all adopt the \tpc\ paradigm~\cite{ruan2019blockchains}, we believe comparing with \tpc\ in \emph{our} experimental setup best illustrates the benefits and trade-offs of \prot.
Existing \tpc\ based approaches primarily focuses on the
UTXO model or other specialized computation models~(ref.~\S\ref{sec:related}), so we need to implement additional support to extend \tpc\ to a generic smart contract model. We next describe implementation details for \tpc\ with generic computation in \S\ref{sec:tpc}.

\subsection{\tpc\ Implementation Details}
\label{sec:tpc}
A \cs\ shard manages all cross-shard 
transactions~\cite{dang2019towards}. We refer to the blockchain maintained
by the \cs\ shard as the \cs\ chain. 
Users send cross-shard transactions along with their potential 
read-write set to the \cs\ shard. The leader of the \cs\ 
shard validates these transactions for read-write conflicts
and, on successful validation, proposes them to be included in 
the \cs\ chain.
Every cross-shard transaction upon its inclusion in the \cs\ 
chain acquires an explicit lock on the set of keys in its read-write 
set. Similar to \prot, the \cs\ shard includes a new cross-shard transaction
only if the transaction does not conflict with any of the pending
cross-shard transactions. 

\Ws\ shards monitor the \cs\ chain for new cross-shard transactions. 
Upon noticing a new cross-shard transaction $\ctx$, involved \ws\ shards commit to
a state of the keys 
mentioned in the read-set of $\ctx$. Each commitment
also carries a proof generated by running consensus within the \ws\
shard. Once the \ws\ proposal is finalized, every replica 
locks the keys mentioned in the read-write set of $\ctx$ from 
any other conflicting transaction until it executes $\ctx$. 
Once commitments from \emph{all} involved shards appear in the
\cs\ chain, these shards follow the same procedure as \prot\ for
data fetching and transaction execution. Upon execution, \ws\ 
shard replicas unlock the keys in $\ctx$ and send an
acknowledgment message to the \cs\ shard -- at this point, the keys become accessible to future intra-shard transactions.

\subsection{Experimental Setup}
\label{sub:expt-setup}


Our experimental setup consists of six \ws\ shards and one \rs\ shard.
Each shard tolerates $f=3$ Byzantine faults.
Thus, each \ws\ shard in \prot\ consists of 7 nodes $(2f+1)$ and the \rs\
shard consists of $10$ nodes $(3f+1)$. Every shard in \tpc\ consists of $10$ nodes 
$(3f+1)$. We run all nodes on 
Amazon Web Services~(AWS) 
t3a.medium virtual machines~(VM) with one node per VM. All VMs have 
2 vCPUs, 4GB RAM, and 5.0 GB/s network bandwidth.
The operating system is Ubuntu 18.04 and the Golang compiler version is 1.13.6.

\two 
\noindent
{\bf Node and network topology.}
We create a overlay network among nodes with 
the following connectivity. Nodes within a shard are pair-wise connected, i.e., form a complete graph. 
In addition, each node is 
connected to $f+1$ randomly chosen nodes from every other shard.
We mimic a setting where each node is placed in one of 10 geographical
locations across different continents.
Instead of placing nodes physically there, we use the measured ping latency~\cite{pingDelay} for every pair of locations and then use the Linux~{\sf tc} tool to insert the corresponding delay to every message. 
We maintain the same network topology and network latency for all our experiments. 

\two
\noindent
{\bf Evaluation methodology.}
We run both \prot\ and \tpc\ for approximately 50 
\rs\ and \cs\ blocks after a initial stabilization period. Every 
\ws\ shard in both \prot\ and \tpc\ generate blocks after every 
$\intv_w=5$ seconds. We vary the block interval of the \rs\ 
chain and \cs\ chain to be $\intv_r=10$.
We test both designs using the benchmark we created in~\S\ref{sec:case},
using Ethereum transactions from 6000 blocks starting at 
block height 7.39 M. This trace comprises of \textasciitilde 14000 cross-shard transactions. To facilitate such evaluation, 
we initialize each shard with the code and state of relevant
smart contracts. 
In all our experiments, we broadcast a new batch of cross-shard 
transactions of fixed size after every \rs\ or \cs\ block. We 
refer to this batch size as the \emph{cross-shard input rate}, 
and test both designs with cross-shard input rates
of 100, 200 and 300. 


\pgfplotsset{small,label style={font=\fontsize{8}{9}\selectfont},legend style={font=\fontsize{7}{8}\selectfont},height=3.8cm,width=1.2\textwidth}

\begin{figure*}[t]
\begin{minipage}[b]{0.48\linewidth}
    \centering
    \pgfplotsset{footnotesize,height=4.5cm, width=0.9\linewidth}
    \begin{tikzpicture}
    \begin{axis}[bar shift=-6pt,
            xlabel={Cross-shard input rate},
            ylabel= {Latency (in seconds)},
            legend columns=2,
            ybar stacked,
            enlargelimits=0.25,
            bar width=10pt,
            symbolic x coords={100,200,300},
            xtick={100, 200, 300},
            ymax=45,
            ymin=6,
        ]
        \addplot [fill=red!30!white] table [x=load, y=s2mean, y error=s2cfi,  col sep=comma] {data/rcCtxLatency.csv};\label{rci}
        \addplot [fill=white] table [x=load, y=s2mean, y error=s2cfi,  col sep=comma] {data/rcpCtxLatency.csv}; \label{rcp} 
    \end{axis}
    \begin{axis}[hide axis, bar shift=6pt,
            ybar stacked,
            enlargelimits=0.25,
            bar width=10pt,
            legend columns=2,
            symbolic x coords={100,200,300},
            xtick={100, 200, 300},
            ymax=50,
            ymin=6,
        ]
        \addlegendimage{/pgfplots/refstyle=rci}\addlegendentry{\prot\ Execution}
        \addlegendimage{/pgfplots/refstyle=rcp}\addlegendentry{\prot\ Wait}
    
        \addplot [fill=blue!30!white] table [x=load, y=s2mean, y error=s2cfi, col sep=comma] {data/tpcCtxLatency.csv}; \label{tpci}
        \addplot [fill=white, postaction={pattern=horizontal lines}] table [x=load, y=s2mean, y error=s2cfi, col sep=comma] {data/tpcpCtxLatency.csv}; \label{tpcp}
        \addlegendentry{2PC Execution}
        \addlegendentry{2PC Wait}
    \end{axis}
    
    \end{tikzpicture}
    \caption{Average confirmation latency of {\em cross-shard} 
    transactions with $\intv_w=5$ seconds and cross-shard input rate of $100,200,$ and $300$.}
    \label{fig:ctx-latency}
\end{minipage}
\hspace{4ex}
\begin{minipage}[b]{0.48\linewidth}
\centering
\pgfplotsset{footnotesize,height=4.5cm, width=0.9\linewidth}
    \begin{tikzpicture}
    \begin{axis}[
        ybar,
        enlargelimits=0.25,
        bar width=10pt,
        /pgfplots/ybar=2pt,
        legend columns=2,
        legend pos=north east,
        xlabel={Cross-shard input rate},
        ylabel= {Latency (in seconds)},
        symbolic x coords={100,200,300},
        grid=minor,
        xtick={100, 200, 300},
        ymax=10,
        ]
        \addplot [fill=red!30!white] [error bars/.cd, y explicit,y dir=both,] table [x=load, y=s2mean, col sep=comma] {data/rcIntLatency.csv}; 
        
        \addplot [fill=blue!30!white] [error bars/.cd, y explicit,y dir=both,] table [x=load, y=s2mean, col sep=comma] {data/tpcItxLatency.csv};
        
        \addlegendentry{\prot}
        \addlegendentry{\tpc}
    \end{axis}
    \end{tikzpicture}
    \caption{Average confirmation latency of {\em intra-shard} 
    transactions with $\intv_w=5$ seconds and 
    cross-shard input rate of $100,200,$ and $300$.}
    \label{fig:itx-latency}
\end{minipage}
\begin{minipage}[b]{0.48\linewidth}
    \centering
    \pgfplotsset{footnotesize,height=4.5cm, width=0.95\linewidth}
    \begin{tikzpicture}
    \begin{axis}[
        ybar,
        enlargelimits=0.25,
        bar width=10pt,
        /pgfplots/ybar=2pt,
        legend columns=2,
        legend pos=north east,
        xlabel={Cross-shard input rate},
        ylabel= {Cross-shard throughput},
        symbolic x coords={100,200,300},
        grid=minor,
        xtick={100, 200, 300},
        ymax=1.1,
        ymin=0.40,
        ]

        \addplot [fill=red!30!white, postaction={pattern=north east lines}] [error bars/.cd, y explicit,y dir=both,] table [x=load, y=s2mean,  col sep=comma] {data/rcoutRate.csv}; 
        \addplot [postaction={pattern=horizontal lines}] [error bars/.cd, y explicit,y dir=both,] table [x=load, y=s2mean, col sep=comma] {data/tpcoutRate.csv};
        
        \addlegendentry{\prot}
        \addlegendentry{\tpc}
    \end{axis}
    \end{tikzpicture}
    \caption{Average cross-shard transaction throughput for $\intv_w=5$ seconds, and cross-shard input rate of 100, 200, and 300.}
    \label{fig:ctx-tput}
\end{minipage}
\hspace{4ex}
\begin{minipage}[b]{0.48\linewidth}
    \centering
    \pgfplotsset{footnotesize,height=4.5cm, width=0.95\linewidth}
    \begin{tikzpicture}
    \begin{axis}[
        ybar,
        enlargelimits=0.25,
        bar width=10pt,
        /pgfplots/ybar=2pt,
        legend columns=2,
        legend pos=north east,
        xlabel={Cross-shard input rate},
        ylabel= {Intra-shard throughput},
        symbolic x coords={100,200,300},
        grid=minor,
        xtick={100, 200, 300},
        ymin=0.40,
        ymax=1.1,
        ]
        \addplot [fill=red!30!white, postaction={pattern=north east lines}] [error bars/.cd, y explicit,y dir=both,] table [x=load, y=s2mean, col sep=comma] {data/rcIntTput.csv}; 
        
        \addplot [postaction={pattern=horizontal lines}] [error bars/.cd, y explicit,y dir=both,] table [x=load, y=s2mean, col sep=comma] {data/tpcIntTput.csv};
        \addlegendentry{\prot}
        \addlegendentry{\tpc}
    \end{axis}
    \end{tikzpicture}
    \caption{Average intra-shard transaction throughput with 
    $\intv_w=5$ seconds and cross-shard input rate of 100, 200, 
    and 300.}
    \label{fig:itx-tput}
\end{minipage}
\end{figure*}

\subsection{Experimental Results}
\label{sub:expt-results}
\noindent
{\bf Confirmation latency.}
Figure~\ref{fig:ctx-latency} gives the average confirmation 
latency of cross-shard transactions in \prot\ and \tpc\ under 
varying cross-shard input rate. 
The confirmation latency of 
a cross-shard transaction is the time elapsed since the first 
time a \rs\ shard replica attempts to include the transaction
in a block till it is executed by one of the
participating \ws\ shard. 
We further divide the latency into two 
parts: {\em wait} and {\em execution} latency. 
The wait latency refers to the time elapsed between the first \emph{attempt} 
to include this transaction in a \rs\ block till it is committed in the \rs\ blockchain. 
The execution latency is the time elapsed 
since the transaction is committed in the \rs\ blockchain till it its execution. 

The wait latency is similar for \prot\ and \tpc. 
But \prot\ has a shorterexecution latency. 
The reason is that the cross-shard execution latency of \prot\ 
only depends on the \ws\ shard block generation interval $\intv_w$.
In contrast, cross-shard execution latency in \tpc\ 
requires at least one additional \rs\ block, i.e, it is 
approximately $\intv_r + \intv_w$.

We then turn to confirmation latency of intra-shard transactions, shown in Figure~\ref{fig:itx-latency}.
In \prot, the latency comes from the fact that \ws\ blocks are finalized only when their commitments are finalized in the \rs\ chain. 
In \tpc, the latency is a result of locking: if an intra-shard transaction needs to read or write a locked account, it is delayed until the lock is released. 
In our benchmark, most intra-shard transactions do not conflict with locked cross-shard assets, so this latency is insignificant. 
After an intra-shard transaction is executed in \tpc\, it is finalized very quickly by the \ws\ shard consensus protocol.
Hence, \prot\ has a worse intra-shard latency than \tpc, and this is the major trade-off in \prot.




\two 
\noindent
{\bf Transaction throughput.}
For a given cross-shard input rate, we measure the transaction 
throughput for cross-shard transactions as the ratio between average
number of cross-shard transactions included per \rs\ block to the
cross-shard input rate. Similarly, for intra-shard transactions 
we measure its throughput as the average (over shards) ratio between 
total number of intra-shard transactions included in the worker 
shards to the total number of intra-shard transactions fired during
the experiment.

Figure~\ref{fig:ctx-tput} illustrates the cross-shard throughput
of \prot\ and \tpc\ under varying cross-shard input rate. 
Recall from~\S\ref{sub:rs-shard}  and~\S\ref{sec:tpc} 
that the \rs\ and the \cs\ chains only include non-conflicting 
transactions in them, and hence not all cross-shard transactions can be immediately included in the \rs\ chain.
In all our experiments of \prot, we observe that the 
cross-shard output rate is greater than 75\% of the cross-shard
input rate whereas cross-shard output rate of \tpc\ is 
approximately 60\% of the corresponding cross-shard input 
rate. The reason is that \tpc\ holds locks on certain accounts for at least one intermediate \cs\ block which results in higher 
conflicts during the inclusion of newer cross-shard transactions. 
On the other hand, conflicts in \prot\ can be resolved prior 
to the next \rs\ block. Lastly, as anticipated, the absolute
value of cross-shard output rate increases linearly with 
increase in cross-shard input rate, as the proposer of 
the \rs\ chain in \prot\ (\cs\ chain in \tpc) has a larger 
number of transactions to choose from for each new \rs\ block. 

Figure~\ref{fig:itx-tput} illustrates the intra-shard throughput
for both \prot\ and \tpc\ for varying cross-shard input rate.
As anticipated, in \prot\ all most all available intra-shard 
transactions are included in every \ws\ block in every shard. 
The slightly less throughput of \tpc\ is due to conflicts between
locked cross-shard transactions and available intra-shard 
transactions. As we have described earlier, since the number
of such conflicting intra-shard transactions is very small in
comparison to the total number of intra-shard transactions, the 
reduction is throughput in \tpc\ is barely noticeable. In 
conclusion, we can say that for our benchmark, both \tpc\ and
\prot\ have almost optimal throughput for intra-shard transactions.


%

\two 
\noindent
{\bf Other findings (not shown).}
In addition to the above results, we observe some other findings that are consistent across our experiments. 
More than 99\% of the state commitments in \prot\ are included in the immediate successor \rs\ block. 
Furthermore, at most one re-organization per shard during the entire duration of the protocol.
The very-low re-organization is a consequence of our dynamic 
re-scheduling of block-proposal time instant in a way that enables 
the future leaders to commit their block with higher probability.
These properties ensure that the intra-shard latency in \prot\ is less than one \rs\ block interval. 
Similarly, almost all commit messages in \tpc\ are also included in the immediate successor \cs\ block. Each node successfully
downloads the data required for a cross-shard transactions within the first two seconds of hearing about the cross-shard transaction.  

\section{Related Work}
\label{sec:related}
\prot\ is partially inspired by the approach of Deterministic Transactions
Execution~(\dte) in distributed databases~\cite{thomson2012calvin}. 
In \dte\ all servers~(shards in our case) first agree on an ordered 
list of transactions and then deterministically execute them in the 
agreed order. Abadi et al.~\cite{abadi2018overview} give a great 
overview of the recent progress and improvements of \dte. 
\dte\ are be made to avoid single points of failure by replicating 
each server across multiple replicas using a crash fault-tolerant 
consensus protocol such as Paxos~\cite{lamport2019part}.
At some level, \prot\ can be viewed as a method to make \dte\ Byzantine fault tolerant.
But \prot\ also differs from fault-tolerant
\dte\ in two major ways: 
First, \prot\ 
tolerates Byzantine failure without using any consensus algorithm 
within the \ws\ shards. 
Second, \dte\ globally orders all the transactions before executing them; in contrast, cross-shard transactions are ordered before being executed whereas intra-shard transactions are optimistically executed before being ordered.

\two
\noindent
{\bf Blockchain sharding.}
Previous blockchain sharding proposals primarily focus 
on increasing the overall throughput of the entire system, with minimal 
emphasize on characterizing and handling cross-shard transactions~\cite{danezis2015centrally,
al2017chainspace,zamani2018rapidchain,wang2019monoxide,dang2019towards,
manuskin2019ostraka}. 
As summarized in~\cite{ruan2019blockchains} almost all prior 
works use minor variants of \tpc\ for cross-shard transactions.

RS-Coin~\cite{danezis2015centrally} and Omniledger~\cite{kokoris2018omniledger} 
are client driven sharded systems in the UTXO model where cross-shard 
transactions are executed using 2PC. RS-Coin is a permissioned system whereas Omniledger 
considers a permissionless model. Chainspace~\cite{al2017chainspace} also uses 
a variant of 2PC for cross-shard transaction where it substitutes the client by a 
inter-shard consensus protocol called S-BAC. 
RapidChain~\cite{zamani2018rapidchain} also considers UTXO based model 
where cross-shard transactions are replaced by dummy transactions at 
every participating shard. These dummy transactions maintain semantic 
properties of the original cross-shard transactions. To execute a 
cross-shard transaction, shards involved in the transaction run 2PC
protocol with every output shard, playing the role of the 2PC transaction 
coordinator and input shards being the server. 

Monoxide~\cite{wang2019monoxide} partitions its participants into 
shards where nodes in each zone run PoW. Monoxide also 
adopts UTXO based data model and runs 2PC for cross-shard transaction. 
Cross shard transactions are executed in the initiator shards and then
the proofs are sent to the receiver shards. Since Monoxide uses PoW, 
the receiver shard needs to wait for a long duration before it
can confidently use the certificates from initiator shard. 
Cross shard transactions in~\cite{dang2019towards} use two-phase
locking~(2PL) and 2PC to achieve atomicity and isolation. To defend against attacks 
from clients who can lock-up shared resources for long periods, they 
replace clients by a distributed committee. They demonstrate that
RapidChain does not achieve atomicity in non-UTXO model.

\two 
\noindent
{\bf State paritioning.}
Our partitioning technique shares similarities with Schism~\cite{schism2010}, a database partitioning system for distributed databases. Schism models the  database as a graph, where a vertex denotes a single record/tuple and an edge connects two records if they are accessed by the same transaction. 
A recent work Optchain~\cite{nguyen2019optchain} improves the placement of 
transaction in a sharded blockchain to reduce the fraction of 
cross-shard transaction. In contrast to our 
graph representation, Optchain models transactions as nodes and 
transaction dependencies as edges. It deals only with the UTXO 
model. It also places more emphasis on temporal balancing, 
where the number of nodes in each shard must be the same at 
all times. 
A concurrent work~\cite{tao2020sharding} focuses on increasing 
throughput by creating individual shards for transactions that solely 
access one particular contract and a single shard for transactions 
that access multiple contracts. 
These works do not address the problem of efficient execution of 
cross-shard transactions.

\two
\noindent
{\bf Sharding and off-chain based solutions.}
Off-chain solutions~\cite{poon2016bitcoin,miller2019sprites,dziembowski2019perun,teutsch2019scalable,kalodner2018arbitrum,das2018yoda} represent an alternative direction to improve blockchain scalability. 
We observe that off-chain solutions and sharding solutions have deep connections.
This is not obvious at all from the current state of the literature partly because the two approaches start out with very different motivations. 
Off-chain solutions shard part of their state/UTXOs among many subset
of $n$ nodes ($n=2$ for payment channels). These nodes process local
transactions, maintain the latest information about 
the assigned state and use the consensus engine, i.e., the blockchain,
to order them globally relative to other shards. 
Recent off-chain
based protocols such 
as~\cite{teutsch2019scalable,kalodner2018arbitrum,das2018yoda} extend the dispute resolution using 
incentives. At their core, sharding schemes  have a similar structure. Typically they use full-fledged consensus within every shard
and some coordination schemes~(so far \tpc) between shards to get rid of the global consensus engine. Our paper deviates from this conventional 
wisdom by removing consensus from the worker; it is thus like a hybrid of 
both sharding and off-chain scalability solutions. 
\section{Conclusion and Future Directions}
\label{sec:discussion}
We have presented \prot, a new paradigm for executing cross-shard
transactions in a sharded system. \prot\ 
has low latency and high throughput for cross-shard transactions in comparison 
with the \tpc\ approach. Also, only the \rs\ shard in \prot\ is 
required to run a consensus protocol; \ws\ shards only vouch for the validity of 
blocks, and hence they require fewer replicas and less communication.



It is plausible to substitute the \rs\ chain with a hierarchy of 
\rs\ chains each coordinating commitments and cross-shard transactions
between a subset of \ws\ shards. Such a hierarchical design 
would allow the system to process more 
cross-shard transactions concurrently. Furthermore, such a design may better exploit 
locality of interaction between different subsets of shards. Extending 
our approach to a hierarchical design is a promising future research direction. 

\section*{Acknowledgments}
The authors would like to thank Amit Agarwal, Jong Chan Lee,
and Zhuolun Xiang for numerous discussions related to the paper. 
The authors would also like to thank the Quorum 
open source community for their responses on queries related to
the Quorum implementation.



\bibliographystyle{ACM-Reference-Format}
\bibliography{references}

\section{Proof of Safety and Liveness}
\label{apx:security}
In this section we will argue about the safety and liveness
of \prot. Informally, safety captures the idea that \prot\ 
ensures a global agreed order on a set of transactions and 
liveness captures that idea that newer transactions are 
continuously included in the global order. Additionally, we
will also prove that for every intra-shard transaction, every 
replica of the corresponding shard executes the intra-shard 
transaction atop a identical starting state. Furthermore, 
for every cross-shard transaction, all the replicas involved
in the cross-shard transaction from all the relevant shard, 
executes the cross-shard transactions atop an identical initial
state. Since, transaction execution are deterministic, this
implies that the final state after executing the cross-shard
transactions are also identical.

Next we will formally define safety and liveness conditions and 
prove that \prot\ ensures the defined safety and liveness. 
\begin{definition} ({\em Safety})
If an honest replica (could be either \ws\ replica or \rs\ replica) 
outputs a block $B$ at height $i$, every other honest replica of 
the same shard will also output block $B$ at height $i$.
\label{def:live}
\end{definition}

\begin{definition} ({\em Liveness})
Transactions~(both cross-shard or intra-shard) sent to honest replicas 
in every shard are included in the blockchain within finite amount of 
time.
\label{def:safe}
\end{definition}

Safety of \prot\ follows directly from the 
safety of the Byzantine fault tolerant consensus protocol used in 
the \rs\ shard. This holds true even during periods of asynchrony
because the partially synchronous consensus algorithm provides safety 
even under asynchrony. 
To elaborate, the consensus algorithm in the \rs\ shard provides global order for all transactions, the intra-shard ones as well as cross-shard ones.
Each transaction is associated with a unique \rs\ block that finalizes it in the \rs\ chain.
Transactions are hence ordered first by their heights in the \rs\ chain, and then by their positions inside the \rs\ block. 

Besides an agreed upon total order, \prot\ also ensures every \ws\ block commitment finalized in the \rs\ chain represents a valid statement. 
The \rs\ shard ensures that at most one \ws\ block at any given 
height from a shard gets finalized and it extends a previously 
finalized \ws\ block in that shard. We next formally prove that this 
implies every honest \ws\ shard replica will have the same sequence 
of \ws\ of blocks in its local blockchain. 

\begin{lemma}
For any given shard $X$, commitment of at most one worker block at
any given height from shard $X$ is included in the \rs\ chain. 
\label{lem:unique}
\end{lemma}
\begin{proof}
For the sake of contradiction, assume that state commitments 
$\cm$ and $\cm'$ of two \ws\ shard blocks $B$ and $B'$, 
respectively, both of at height $i$ was included in the 
reference chain. Without loss of generality, let $\cm$ be 
is included before $\cm'$ and let $P'$ be the \rs\ block 
that includes $\cm'$.
Then, according to Algorithm~\ref{algo:ref-valid}, no honest
replica will vote for $P'$. But, since $P'$ is a committed 
\rs\ block, this implies that at least $f+1$ honest replica
voted for $P'$. Hence we get a contradiction. 
\end{proof}

We next argue that when two honest replica in \ws\ shard outputs
their local chain, chain on one replica will be a prefix of the 
the chain output by the other replica. For two chain~(blockchain)
$A,B$, we use $A\px B$ to denote that the chain $A$ is a prefix 
of chain $B$. Hence, for \prot,
\begin{theorem}
Let $n,n'$ be the two arbitrary honest replica of a given
shard $X$, and let $\ch$, $\ch'$ be their respective local 
chain. Then either $\ch \px \ch'$ or $\ch' \px \ch$.
\label{thm:prefix}
\end{theorem}
\begin{proof}
For the sake of contraction assume that $\ch \not\px \ch'$ 
and $\ch' \not\px \ch$. Also, let $B^* \in \ch \cap \ch'$ be 
the last block where both the chain agree. Also, let $B,B'$
be the latest committed block of $\ch,\ch'$ respectively. 
Observe that $\hgt(B) > \hgt(B^*)$ and $\hgt(B') >\hgt(B^*)$.
Hence, block $B$ does not extend $B'$ and vice versa. 

From Lemma~\ref{lem:unique}, we know that both block $B$ and 
$B'$ have distinct height. Without loss of generality, let 
the height of $B$ be smaller than height of $B'$. Then if 
commitment of block $B$ appears before commitment of $B'$, 
and if $P'$ is the \rs\ block that includes commitment of 
$B'$; then since, $B'$ does not extend $B$, no honest \rs\ 
replica will vote for validity of $P'$. However, since $P'$ 
is a committed \rs\ block, this implies that at least 
$f+1$ honest replicas indeed voted for $P'$ resulting in a 
contradiction. 

Alternatively, if a commitment of $B$ appears before commitment
of $B'$, then no honest replica will vote for a \rs\ block that
includes commitment of $B'$. Hence, we again get a contradiction
by the same argument as above. 
\end{proof}

We will next argue about liveness of \prot\ during periods of 
synchrony. As in safety, the liveness of the \rs\ shard follows 
directly from the liveness guarantee of the underlying Byzantine
Fault tolerant consensus protocol. Hence, we will focus on the
liveness of \ws\ shards. In particular, we will prove the 
liveness of \ws\ in two steps. First, we will show that during 
periods of synchrony, if the leader, say $L$, of worker shard in 
view $v$ is honest, and reference shard also has a honest 
leader during the same time interval, then  $L$ will be able
to successfully commit a new \ws\ block on to the \rs\ chain. 
Next, we will argue that \prot\ such periods where both the 
leader of \rs\ shard and \ws\ shard are simultaneously honest,
occur infinitely often. 

Let $n_r$ be the number of replicas in the \rs\ shard and $f_r$ be 
the maximum number of Byzantine replicas in the \rs\ shard. 
Similarly, let $n_w$ and $f_w$ be the number of replicas and 
maximum number of Byzantine replicas in a worker shard 
respectively. Recall $n_r = 3f_r + 1$ and $n_w = 2f_w+1$. 
Also, let $\alpha$ be the ratio between rate of change of views
of a worker shard $X$ and the reference shard. For example, 
$\alpha=1$ implies that view change in both $X$ and the \rs\ 
chain happens at the same speed. Similarly, $\alpha=2$ 
implies that view change in worker shard $X$ happens twice
faster than the view change of \rs\ shard.

\begin{claim}
Let $v,v' \ge v + n_r$ be any two distinct views in \rs\ shard. 
Then, for every \ws\ shard $X$, $n_r \ge n_w$, and $\alpha \ge 1$, 
there exists a honest view $v^*$ between view $v$ and $v'$, i.e., 
view with a honest \rs\ leader $L_r$, such that during view $v^*$, 
an honest replica $L_w$ of $X$ becomes the leader of $X$. Here on,
we refer to such view as HH view. 
\label{clm:event}
\end{claim}
\begin{proof}
Let $v' = v + n_r$. Then, during view $v$ and $v'$, exactly 
$\alpha n_r$ nodes will become leader of $X$. Moreover, 
at least $\alpha(n_r -f_r)$ \ws\ leaders will co-exists with
honest \rs\ shard leaders. Also, among $\alpha n_r$ leaders
at most $f_w \lfloor\frac{\alpha n_r}{n_w}\rfloor + f_w$
them will by Byzantine. 

Thus $\alpha(n_r -f_r) >f_w \lfloor\frac{\alpha n_r}{n_w}\rfloor+f_w$
will imply that at least one honest \ws\ leader will co-exist with
honest \rs\ leader. Hence, solving the inequality, the condition 
we get:
\begin{align}
    \alpha(n_r -f_r) &> f_w \lfloor\frac{\alpha n_r}{n_w}\rfloor+f_w \notag \\ 
    \alpha(n_r -f_r) & \ge f_w \frac{\alpha n_r+ 1}{n_w} \notag \\ 
    \alpha(n_r -f_r) & \ge \frac{1}{2} (\alpha n_r+ 1) \notag \\ 
    \alpha(2f_r+1) & \ge \frac{1}{2} (\alpha (3f_r+1)+1) \notag \\ 
    \alpha f_r + \alpha - 1 & \ge 0 \label{eq:live}
\end{align}

For $\alpha \ge 1$, equation~\ref{eq:live} is is always true. 
\end{proof}

An immediate corollary of the above claim is that in an infinite
execution of the protocol, such  honest reference leader and 
honest worker leader replica pairs will occur infinitely often. We 
will next use Claim~\ref{clm:event} to prove liveness of 
\prot. 

\begin{theorem}
During period of synchrony, during a HH view, the corresponding 
leader of \ws\ shard $X$, can successfully commit a new \ws\ 
block in the \rs\ chain. 
\label{thm:live}
\end{theorem}
\begin{proof}
During HH, let $v_w,v_r$ be the view of \ws\ and \rs\ shard respectively.
Let $L_w$ and $L_r$ be the corresponding honest leader for view $v_w$ 
and respectively. Also, let $t_r$ be the time instant $L_r$ enters the
view $v_r$. Then by time $t_r + \Delta$, $L_r$ will receive all 
messages sent in view $v_r-1$. Then by time $t_r + 2\Delta$, $L_w$ 
(in fact all worker shard replicas) will receive the latest \rs\ 
block. Since $L_w$ is connected to $f_w+1$ replicas from each 
shard, by time $t_r + 4\Delta$ have the data required to execute
cross-shard transactions up to the latest \rs\ block. 
Then, if at time $t_r + 4\Delta$, if $L_w$ proposes a new
 block $B$, by $t_r + 6\Delta$, $L_w$ will have a certificate 
for $B$. If at time $t_r + 6\Delta$, $L_w$ forwards the block to 
$f_r+1$ \rs\ replicas, by time $t_r + 8\Delta$, $L_r$ will know 
about commitment of $B$. Hence, if $L_r$ proposes its block after
time $t_r + 8\Delta$, the proposed block will include the 
commitment of $B$. And hence by the liveness guarantee of underlying
Byzantine Fault Tolerant consensus protocol of \rs\ chain, commitment 
of $B$ will be finalized. This implies \ws\ block $B$ be finalized.
\end{proof}

\begin{algorithm*}
    \caption{\Rs\ shard block creation at replica (leader) $L$ in view $v$}
    \label{algo:ref-propose}
    \begin{algorithmic}[1]
        \LineComment{{\em Inputs to the block proposal algorithm}}
        \State  $B_{\ell^{(i)}}:$ latest committed block of shard $X_{i}$ for each shard.
        \State $P_{s^{(i)}}:$ Reference block mentioned in  $B_{\ell^{(i)}}$ for each $i$
        \State{\bf $P_{l}:$} Latest \rs\ block known to $L$
        \State{\bf $P_{l+1}:$} Block $L$ wants to propose
        \State \LineComment{{\em We will refer to the cross-shard transactions that 
        appear after the last commit from shard $X_i$ as the pending cross-shard 
        transaction of $X_i$ and denote them by $Q_i$}.}
        \State $Q_i:$ cross-shard transactions that appear after $P^{(i)}_s$ and
        involves $X_i$.
        \State $\br^{(i)}$: Keys that cross-shard transactions in $Q_i$ intends to read from.
        \State $\bw^{(i)}:$ Keys that cross-shard transactions in $Q_i$ intends
        to write to.
        \State
        \State $new_\ctx:$ New cross shard transactions available at $L$
        \State $new_\cm:$ New state commitment transactions available at $L$
        \State 
        \LineComment{Pick set of valid state commitments $S_{l+1}$ from the available state commitments}
        \State $S_{l+1} \gets \emptyset $
        \For {each shard $X_i$}
            \State $valid_{X_i} = \emptyset $ \Comment{Set of new valid state commitments from shard $X_i$.}
            \For {each state commitment $\cm = \la B_j,\bh_j \ra$ from $X_i$ in $new_\cm$} 
                \If {$B_j$ extends $B^{(i)}_\ell$ \& $B_j$ is certified \& $\bq_{s^{(i)},l}$ is empty }
                    \State $valid_{X_{i}} \gets valid_{X_{i}} \cup \{\cm_j\}$
                \EndIf
            \EndFor
            \State $\cm^*_i \gets \max\{valid_{X_i}\}$ \Comment{Pick the commitment with longest hash chain, i.e., highest $|\bh_j|$}
            \State $S_{l+1} \gets S_{l+1}  \cup \{\cm^*_i\}$
            \State $Q_i \gets \emptyset $; $\bw^{(i)} \gets \emptyset$; $\br^{(i)} \gets \emptyset$  \Comment{Resetting cross-shard transactions and the associated read-write sets.}
        \EndFor 
        \State
        \LineComment{Pick set of new cross-shard transactions $C_{l+1}$ from the available cross-shard transactions}
        \State $C_{l+1} \gets \emptyset $ 
        \For{each $\ctx \in new_\ctx$}
            \State Let $X_{\ctx}$ be the set of shards required for executing $\ctx$
            \For{${X_j \in X_\ctx}$} 
                \State Let $R^{(j)}_\ctx, W^{(j)}_\ctx$ be the read and write set of $\ctx$ in $X_j$, respecitvely.
                \If {$R^{(j)}_\ctx \cap \bw^{(j)} = \emptyset $}
                    \State $C_{l+1} \gets C_{l+1} \cup \{\ctx\}$; $\bw^{(j)} \gets \bw^{(j)} \cup W^{(j)}_\ctx$; $\br^{(j)} \gets \br^{(j)} \cup R^{(j)}_\ctx$ \Comment{Update pending cross-shard transactions}
                \EndIf
            \EndFor
        \EndFor
        \State
        \State $P_{l+1} \gets \la l+1, \hs(P_l), S_{l+1}, C_{l+1} \ra$
    \end{algorithmic}
\end{algorithm*}

\begin{algorithm*}
    \caption{Validation of block $ P_{r} = \la r, \hs(P_{r-1}),S_r,C_r \ra $ at replica $n$ of reference shard}
    \label{algo:ref-valid}
    \begin{algorithmic}[1]
        \LineComment{{\em Input to the block validation algorithm. For simplicity, let $P_{r-1}$ is the latest reference block known to $n$.}}
        \State  $B_{\ell^{(i)}}:$ latest committed block of shard $X_{i}$ for each shard up to $P_{r-1}$ .
        \State $P_{s^{(i)}}:$ Reference block mentioned in  $B_{\ell^{(i)}}$ for each $i$
        \State
        \State $Q_i:$ cross-shard transactions that appear after $P_{s^{(i)}}$ and
        involves $X_i$.
        \State $\br^{(i)}$: Keys that cross-shard transactions in $Q_i$ intends to read from.
        \State $\bw^{(i)}:$ Keys that cross-shard transactions in $Q_i$ intends
        to write to.
        \State
        \For {every $\cm_j = \la B_j, \bh_j\ra \in S_r$}
            \If {$\lnot$($B_j$ extends $B_{\ell^{(i)}}$ \& $B_j$ is certified \& $\bq_{s^{(i)},l}$ is empty)}
                \State Output {\bf invalid} $P_r$; {\bf return}
            \EndIf
        \EndFor
        \State
        \For {every $\cm_j = \la B_j, \bh_j\ra \in S_r$}
            \State Let $X_j$ be the worker shard that commits $\cm_j$
            \State $Q_j \gets \emptyset $; $\bw^{(i)} \gets \emptyset$; $\br^{(i)} \gets \emptyset$ \Comment{Temporarily set values for $Q_i, \bw^{(i)},$ and $\br^{(i)}$ to be empty sets}
        \EndFor
        \State
        \For{each $\ctx \in C_r$}
            \State Let $X_{\ctx}$ be the set of shards required for executing $\ctx$
            \For{${X_j \in X_\ctx}$} 
                \State Let $R^{(j)}_\ctx, W^{(j)}_\ctx$ be the read and write set of $\ctx$ in $X_j$, respectively.
                \If {$ R^{(j)}_\ctx \cap \bw^{(j)} \ne \emptyset$}
                    \State Output {\bf invalid} $P_r$; Reset $Q_i, \bw^{(i)},$ and $\br^{(i)}$ to original state $\forall i \in [k]$
                    \State {\bf return}
                \Else
                    \State $Q_j \gets Q_j \cup \{\ctx\}$; $\bw^{(j)} \gets \bw^{(j)} \cup W^{(j)}_\ctx$; $\br^{(j)} \gets \br^{(j)} \cup R^{(j)}_\ctx$
                \EndIf
            \EndFor
        \EndFor
        \State
        \State Output {\bf valid} $P_r$.
    \end{algorithmic}
\end{algorithm*}

\begin{algorithm*}
    \caption{Block creation at replica~(leader) $L$ of a \ws\ shard $X$}
    \label{algo:ws-propose}
    \begin{algorithmic}[1]
        \State {\bf Inputs.}
        \State $B_i:$ Latest committed block of $X$ with state $\st_i$
        \State $P_i:$ \Rs\ block at height $r_i$ that includes $\cm_i = \la \st_i,\ldots \ra$
        \State $B_j:$ Latest \ws\ block of $X$ (may be $B_i = B_j$)
        \State $P_j:$ Latest \rs\ block at height $r_j$ known to $L$
        \State $\bq_{i,j}:$ Cross-shard transactions involving $X$ that appears between \rs\ block $P_i$ and $P_j$ (both inclusive).
        \State $\bt_{i,j}:$ Intra-shard transaction between \ws\ 
        block $B_i$ and $B_j$ (both inclusive).
        \State $T_{j+1}:$ Newly available intra-shard transaction.
        \State
        \If {$\bq_{i,j}$ is not {\em empty} even when $(i=j)$}
            \State parent block $:= B_i$
            \State pick $\st_i$ as the starting state
            \State execute $\bq_{i,j}$ and $\bt_{i,j+1} = \bt_{i,j}\cup T_{j+1}$ atop $\st_i$, i.e., $\st_{j+1} \gets \Pi(\bt_{i,j+1}, \Pi(\st_i, \bq_{i,j}))$
            \State $B_{j+1} \gets \la \hs(B_i), \st_{j+1}, \bt_{i,j+1}, r_i,\hs(P_i) \ra$
        \Else
            \State parent block $:= B_j$
            \State pick $\st_j$ as the starting state
            \State execute $T_{j+1}$ atop $\st_j$, i.e., $\st_{j+1} \gets \Pi(\st_j, T_{j+1})$.
            \State $B_{j+1} \gets \la \hs(B_j), \st_{j+1}, T_{j+1}, r_j,\hs(P_j) \ra$
        \EndIf
    \end{algorithmic}
\end{algorithm*}

\begin{algorithm*}
    \caption{Validation of block $B_j = \la\hs(B_{j'}), \st_j, T_j, r_{j},\hs(P_{j}) \ra$ at replica $n$ of a \ws\ shard $X$}
    \label{algo:ws-valid}
    \begin{algorithmic}[1]
        \State {\bf Inputs.}
        \State $B_j:$ Newly proposed block at $j$.
        \State $B_{j'}:$ Parent of block $B_j$ with state $\st_{j'}$ .
        \State $P_j:$ Reference block mentioned in $B_j$ at height $r_{j}$.
        \State $T_j:$  Intra-shard transactions included in $B_{j}$.
        \State $B_l:$ Latest committed block of $X$ with state $\st_l$ known to $n$.
        \State $P_l:$ \Rs\ block at height $r_i$ that includes $\cm_l = \la \st_l,\ldots \ra$
        \State $P_k:$ Latest \rs\ block at height $r_k$ known to $n$
        \State $\bq_{j',j}:$ Cross-shard transactions involving $X$ 
        included in \rs\ blocks after $B_{j'}$ and up to $P_j$, if any.
        \State $\bq_{j+1, k}:$ Cross-shard transactions involving $X$ 
        included in \rs\ block after $P_j$ and up to $P_k$, if any.
        \State $\bq_{l, j}:$ Cross-shard transactions involving $X$ 
        included in \rs\ block from $P_l$  up to $P_j$, if any.
        \State
        \State $\st_j \gets \perp$  
        \If {$B_j$ extends $B_l$ \& $\hgt(P_k) \ge \hgt(P_j)$ \& $\bq_{j+1,k}$ is empty}
        \Comment{extends means "is a descendant of"}
            \LineComment{{\em Here $\hgt(P)$ refers to the height of the block; also $\bq_{a,b}$ is empty if $\hgt(P_a) > \hgt(P_b)$}}
            \If {$B_j$ = $B_l$}
                \State $\st'_{j} \gets \Pi(\Pi(\st_{j'},\bq_{l,j}),T_j)$
            \ElsIf {$\hgt(B_{j'}) > \hgt(B_l)$ \& $B_{j'}$ is certified \& $\bq_{j',j}$ is empty} 
                \State $\st'_{j} \gets \Pi(\st_{j'},T_j)$
            \EndIf
        \EndIf
        \State
        \If {$\st'_j \ne \perp$ and $\st'_j = \st_j$}
            \State sign; {\bf return}
        \EndIf
        \State do not sign; {\bf return}
    \end{algorithmic}
\end{algorithm*}

            


\end{document}